\documentclass[showpacs,aps,prb,reprint,superscriptaddress,floatfix]{revtex4-2}
\usepackage{bm}
\usepackage{graphicx}
\usepackage{graphics}
\usepackage{mathtools}
\usepackage{amsmath}
\usepackage{amsfonts}
\usepackage{amssymb}
\usepackage{epstopdf}
\usepackage{hyperref}
\usepackage{hyperref}
\hypersetup{
    colorlinks,%
    citecolor=blue,%
    linkcolor=blue,%
    urlcolor=blue
}
\usepackage{tikz}\usetikzlibrary{petri}
\usepackage{soul}

\begin{document}

\newcommand{\be}   {\begin{equation}}
\newcommand{\ee}   {\end{equation}}
\newcommand{\ba}   {\begin{eqnarray}}
\newcommand{\ea}   {\end{eqnarray}}
\newcommand{\ve}  {\varepsilon}

%%%%%%%%%%%%%%%%%%%%%%%%%%%%%%%%%%%%%%%%%%%%%%%%%%%%%%%%%%%%%%%%%%%%%%%%%%%%%%%
\title{ Superconductivity from spin fluctuations and long-range interactions in magic-angle twisted bilayer graphene }

\author{Lauro B. Braz}
\affiliation{
Instituto de F\'{\i}sica, Universidade de S\~ao Paulo, Rua do Mat\~ao 1371, S\~ao Paulo, S\~ao Paulo 05508-090, Brazil
}
\affiliation{
 Department of Physics and Astronomy, Uppsala University, Box 516, S-751 20 Uppsala, Sweden
}
\author{George B. Martins}
\affiliation{Instituto de F\'isica, Universidade Federal de Uberl\^andia, 
Uberl\^andia, Minas Gerais 38400-902, Brazil.}
\author{Luis G.~G.~V. Dias da Silva}
\affiliation{
Instituto de F\'{\i}sica, Universidade de S\~ao Paulo, Rua do Mat\~ao 1371, S\~ao Paulo, S\~ao Paulo 05508-090, Brazil
}

\date{ \today }

\begin{abstract}

Magic-angle twisted bilayer graphene (MATBG) has been extensively explored both theoretically and experimentally as a suitable platform for a rich and tunable phase diagram that includes ferromagnetism, charge order, broken symmetries, and unconventional superconductivity. In this work, we investigate the intricate interplay between long-range electron-electron interactions, spin fluctuations, and superconductivity in MATBG. By employing a low-energy model for MATBG that captures the correct shape of the flat bands, we explore the effects of short- and long-range interactions on spin fluctuations and their impact on the superconducting (SC) pairing vertex in the matrix Random Phase Approximation (matrix RPA). We find that the SC state is notably influenced by the strength of long-range Coulomb interactions. Interestingly, our matrix-RPA calculations indicate that there is a regime where the system can traverse from a magnetic phase to the SC phase by
\emph{increasing} the relative strength of long-range interactions compared to the on-site ones.
These findings underscore the relevance of electron-electron interactions in shaping the intriguing properties of MATBG and offer a pathway for designing and controlling its SC phase.

\end{abstract} 
% \pacs{ APS does not use it anymore}
%\keywords{Quantum Spin-Hall effect, Edge transport, Topological insulators}

\maketitle

\section{Introduction}
\label{sec:Intro}

Magic-angle twisted bilayer graphene (MATBG) has  been a prominent research topic in materials science
due to its highly tunable phase diagram, which displays similarities to the phase diagram of the cuprate superconductors~\cite{caoUnconventionalSuperconductivityMagicangle2018,polshynLargeLinearintemperatureResistivity2019,caoStrangeMetalMagicAngle2020}. Experimentally, it was found ferromagnetism at half-filling~\cite{sharpeEmergentFerromagnetismThreequarters2019,serlinIntrinsicQuantizedAnomalous2020}, charge order at quarter filling~\cite{jiangChargeOrderBroken2019}, broken symmetry orders at half-integer fillings~\cite{bhowmikBrokensymmetryStatesHalfinteger2022}, evidence for strong correlations~\cite{choiElectronicCorrelationsTwisted2019,balentsSuperconductivityStrongCorrelations2020}, including competing orders \cite{caoNematicityCompetingOrders2021} and Chern insulator states~\cite{nuckollsStronglyCorrelatedChern2020}, as well as corroboration for unconventional superconductivity \cite{caoUnconventionalSuperconductivityMagicangle2018,caoNematicityCompetingOrders2021}. Nonetheless, there is a lack of consensus regarding an intrinsic superconducting (SC) pairing mechanism consistent with the observed plethora of interesting phases in this system. 

Interacting models of MATBG have the challenge of conciliating the magnetic and SC phase scenarios observed in the system.
This interplay between magnetism and superconductivity corroborates the thesis that electron-electron (e-e) interactions play a relevant role in the MATBG phase diagram, establishing the relevance of magnetic 
fluctuations \cite{Roy:Phys.Rev.B:121407:2019}, in analogy with other systems such as the iron pnictides~\cite{Scalapino2012}, 
where magnetic fluctuations dominate the (undoped) normal state and may cause a SC gap to 
emerge~\cite{graserNeardegeneracySeveralPairing2009}. 

On the theory side, Hubbard-like models for MATBG with interaction-hopping ratios $U/t \sim 1$ have been proposed \cite{kerelskyMaximizedElectronInteractions2019}, which suggest that e-e interaction effects can be strong and thus spin and charge fluctuations could have relevant contributions to the origin of the SC phase in MATBG. 
The role of e-e interactions in the insulating and SC phases of MATBG was investigated experimentally in Ref.~\cite{liuTuningElectronCorrelation2021} by effectively tuning the charge screening using a Bernal bilayer near the MATBG sample. 
The results suggest that a larger screening (weaker Coulomb interaction) tends to reduce the insulating gap in MATBG, making the insulating states less robust.
On the other hand, the SC critical temperature at optimal doping tends to \emph{increase} for larger screening as compared to small screening, effectively enhancing the stability of the SC phase. 
The understanding of the microscopic picture behind these findings might shed light on the conceptual dispute of e-e versus electron-phonon mechanisms proposed for superconductivity in MATBG~\cite{Lian2019}. 

In this paper, we show that
the shape of the SC phase 
is strongly influenced not only by local but, interestingly, by \emph{long-range} interactions 
\footnote{In this paper, we refer to Hubbard-like interactions beyond nearest-neighbors as ``long-range'', although we are aware that some authors, e.g. Ref.~\cite{defenuLongrangeInteractingQuantum2023}, prefer to use the term ``non-local'' for such terms. }. 
The normal state is modeled by the low-energy 2-orbital model parameterized in a tight-binding (TB) 
Hamiltonian \cite{koshinoMaximallyLocalizedWannier2018} that reproduces the flat band dispersion of MATBG at the ``magic angle'' $\theta \approx 1.05^\text{o}$ captured by 
the continuum model of Ref.~\cite{bistritzerMoireBandsTwisted2011}.
In addition, this non-interacting model accounts 
for the correct Fermi surface topology of the system at the experimentally relevant band fillings~\cite{koshinoMaximallyLocalizedWannier2018,lucignanoCrucialRoleAtomic2019a}. 

We then employ matrix Random Phase Approximation (matrix RPA) calculations \cite{esirgenMathitWavePairing1999a, graserNeardegeneracySeveralPairing2009, altmeyerRoleVertexCorrections2016,wuIdentificationSuperconductingPairing2019} to investigate the effect of short- and long-range 
interactions on the spin fluctuations and their influence on 
the SC pairing vertex and, consequently, on the onset of the SC phase. 
We focus our attention on the dependence of the SC 
state as a function of both the band-filling factor and long-range Coulomb interaction strength.

Our results show that, as a general trend, both on-site 
and long-range interactions favor an SC ground state up to 
a maximum interaction strength beyond which a Stoner-like 
magnetic instability sets in. More interestingly, 
for some values of the band-filling factor, 
there is a region in the phase diagram where the (spin) Stoner boundary 
shows a ``duckbill'' shape. As a consequence, 
in this region, the SC phase can 
be  enhanced by \emph{decreasing} the long-range interactions relative 
to the local ones, until eventually crossing the Stoner boundary, 
into the magnetic phase, from above. 
In fact, also as a consequence of this 
duckbill shape, there is a small region of the parameter space where one can 
even tune the system out of the magnetic (and into the SC phase, i.e., across the Stoner boundary) 
by \emph{increasing} the strength of long-range interactions relative to on-site (local) ones.

Our RPA calculations show that such 
``reentrant behavior'' [moving from the square to the circle 
in Fig.~\ref{fig:lambda_filling}(b)] is linked to a sharp increase in the spin-singlet pairing vertex at
finite momenta [compare Figs.~\ref{fig:RPA_chi}(b) and (d)]. 
These results are consistent with a scenario in which e-e correlations, 
especially long-range ones, can play an important role in the
pairing mechanism in  MATBG, as the interplay of band topology and interactions can lead to the strengthening of the SC state in flat band systems \cite{tormaSuperconductivitySuperfluidityQuantum2022a}.

This paper is organized as follows: the microscopic model and details of the matrix-RPA calculations are given in Sec.~\ref{sec:modelmethods}, while the main results for the SC pairing vertex are given in \ref{sec:pairingvertex}. One of our main results, the appearance of a ``duckbill''-shaped feature in the Stoner boundary line, is discussed in Sec.~\ref{sec:duckbill}. Our overall conclusions are summarized in Sec.~\ref{sec:Conclusions}.

%%%%%%%%%%%%%%%%%%%%%%%%%%%%%%%%%%%%%%%%%%%%%%%%%%%%%%%%%%%%%%%%%%%%%%%%%%%%%%%
\section{Model and Methods}
\label{sec:modelmethods}

\subsection{Effective Hamiltonian}
\label{sec:effectiveHami}

The formulation of effective low-energy models for MATBG is a challenging task. 
Near the magic twist angle $\theta \approx 1.05$°, MATBG presents a very large unit cell with lattice constant $L_M \approx 13.4$ nm and, conversely, the momentum-space Brillouin Zone (BZ) of the superlattice is very small \cite{LopesdosSantos:Phys.Rev.Lett.:256802:2007,bistritzerMoireBandsTwisted2011}, making atomistic-type real-space tight-binding descriptions a computationally costly endeavor \cite{lothmanNematicSuperconductivityMagicangle2022}. 

A more sought-after approach is to build well-localized Wannier orbitals describing the low-energy flat bands which have been shown to be effectively detached from the conduction and valence bands for $\theta \approx 1.05$°~\cite{bistritzerMoireBandsTwisted2011}. Early proposals for a tight-binding two-orbital model based on optimized Wannier functions \cite{kangSymmetryMaximallyLocalized2018,koshinoMaximallyLocalizedWannier2018} were able to successfully describe the band dispersion of the flat bands, provided that the Wannier orbitals are greatly optimized, leading to the need to include very long-range hopping terms \cite{koshinoMaximallyLocalizedWannier2018}.  Corrugation (i.e. vertical relaxation) effects are treated effectively in this model, which correctly accounts for the DFT band structure at the magic angle \cite{lucignanoCrucialRoleAtomic2019a} and shows agreement with fully relaxed continuum models \cite{Carr:Phys.Rev.Res.:013001:2019}.

As it was later realized, such models are subject to the so-called ``topological obstruction'' \cite{Zou:Phys.Rev.B:085435:2018,Po:Phys.Rev.X:031089:2018,poFaithfulTightbindingModels2019a,songAllMagicAngles2019}, in which some of the symmetries of the continuum model (including emergent ones) cannot be captured by the effective tight-binding model. Alternative formulations, involving six, eight, or even ten orbitals per valley and per spin,  have been proposed \cite{poFaithfulTightbindingModels2019a,Carr:Phys.Rev.Res.:013001:2019,calderonInteractions8orbitalModel2020}. 

We recall that an important factor in the matrix-RPA description of the (interacting) magnetic susceptibility is the accuracy of the non-interacting bands and the shape of the Fermi surface (FS) for a given filling as it leads to strong nesting effects  \cite{graserNeardegeneracySeveralPairing2009,wuHarmonicFingerprintUnconventional2020}. As such, while the topological obstruction in the two-orbital model is a well-known issue, we are interested in the low-energy phenomena leading to superconductivity in the system. Such regime will be dominated by the flat bands, and the choice of a two-orbital model \cite{kangSymmetryMaximallyLocalized2018} yielding a good enough description of the band dispersion is an acceptable compromise for the goals of the present work.

\begin{figure}[t]
\begin{center}
\includegraphics[width=1.0\columnwidth]{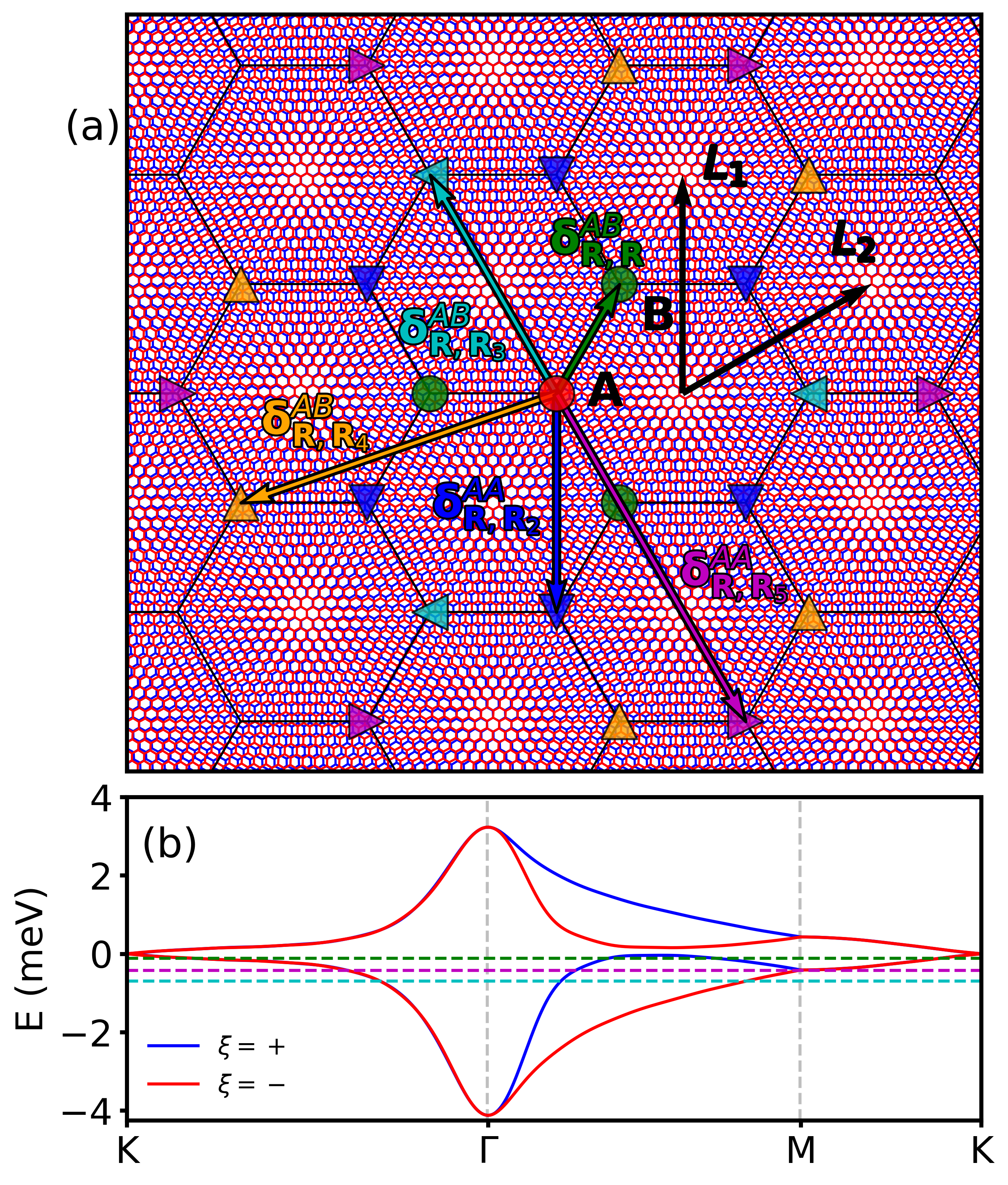}
\caption{Schematic of the non-interacting model we used, 
	obtained by Wannierizing the TBG continuum model~\cite{koshinoMaximallyLocalizedWannier2018}. 
	Panel (a) shows a depiction of the real-space moir\'e superlattice for $\theta=3.2^o$  centered at an $A$ sublattice site 
	(red circle) with first neighbor $B$ sublattice sites (green circles) and up to fifth nearest-neighbor sites (green circle, blue, cyan, orange, and magenta triangles). Also shown are the lattice vectors $\boldsymbol{L}_1$ and $\boldsymbol{L}_2$ and some of the $\boldsymbol{\delta}^{p p^\prime}_{\boldsymbol{R},\boldsymbol{R}^\prime}$ vectors defined in Eq.~\eqref{eq:Uq} for $\boldsymbol{R}^\prime \equiv \boldsymbol{R}_n=\boldsymbol{R}+ n_1 \boldsymbol{L}_1 + n_2 \boldsymbol{L}_2$ with $n_1$ and $n_2$ integers. 
 In panel (b) 
 %and (c) \LUIS{Do we really need panel (c)?}, 
 we present the  TBG band structure
 %and density of states (DOS), respectively, 
 obtained 	with the TB model used here. The red and 
	blue lines in panel (b) denote the $\xi=\pm$ orbitals. The horizontal dashed lines in panel (b) denote the doping levels of the Fermi surfaces shown in Fig.~\ref{fig:lambda_filling}: $\nu=-0.81$ (green), $\nu=-2.48$ (magenta), and $\nu=-3.05$ (cyan).
}
\label{fig:model}
\end{center}
\end{figure}

As such, we adopt the following model Hamiltonian 
\begin{equation}
    H = H_0 + H_\text{int},
\end{equation} 
where the non-interacting term $H_0$ is given by the two-band model of Ref.~\cite{koshinoMaximallyLocalizedWannier2018}:
\begin{equation}
    H_0 = \sum_{\boldsymbol{R}, \boldsymbol{R}^\prime  }\sum_{p p^\prime \sigma} \sum_{\xi} t^{p p^\prime}_{\boldsymbol{R},\boldsymbol{R}^\prime}e^{i\xi\phi^{p p^\prime}_{\boldsymbol{R},\boldsymbol{R}^\prime}}c_{\boldsymbol{R} p \xi \sigma}^{\dagger}c_{\boldsymbol{R}^\prime p^\prime \xi \sigma} \; ,
    \label{eq:TBmodel}
\end{equation} 
where $c_{\boldsymbol{R} p \xi \sigma}^{\dagger}$ creates an 
electron with spin $\sigma=\{\uparrow,\downarrow\}$, valley 
index $\xi=\pm$, in the Wannier state $|\boldsymbol{R}, p \rangle$ 
centered at position $p=A,B$ [see Fig.~\ref{fig:model}(a)], in the unit cell located at the 
moir\'e lattice vector $\boldsymbol{R}$. Notice that $H_0$ is 
block-diagonal in the valley index $\xi$. 

In our calculations, we followed Ref.~\cite{koshinoMaximallyLocalizedWannier2018} and considered all hoppings $t^{p p^\prime}_{\boldsymbol{R},\boldsymbol{R}^\prime}$ and phases $\phi^{p p^\prime}_{\boldsymbol{R},\boldsymbol{R}^\prime}$ connecting sites at distances  $r \lesssim 9 L_M$ from each other. As shown in Fig.~\ref{fig:model}, the resulting two-band dispersion reproduces well the flat bands from the continuum model~\cite{bistritzerMoireBandsTwisted2011} at the magic angle $\theta=1.05^o$ . 

We show the resulting band structure and DOS in Figs.~\ref{fig:model}(b)-(c). As usual, one can define the band filling factor $\nu = 4(n/n_s) \in [-4,4]$ where $n$ is the carrier density and $n_s = 4/A'$ ($A'$ unit cell area) is the superlattice carrier density. The filling factor essentially counts the number of extra electrons and holes per superlattice area, with the limiting cases $\nu\!=\!+(-)4$ corresponding to a full (empty) band, while $\nu\!=\!0$ corresponds to charge neutrality. 

%-----------F I G U R E  1 ------
%-----------E N D  F I G U R E  1 ------

The interacting part of the Hamiltonian, $H_\text{int}$, is given by 
\begin{equation}
H_{\text{int}} = \sum_{\boldsymbol{R}, \boldsymbol{R}^\prime  }\sum_{p p ^\prime \sigma\bar{\sigma}} U^{p p^\prime}_{\boldsymbol{R},\boldsymbol{R}^\prime} \hat{N}_{\boldsymbol{R} p  \sigma} \hat{N}_{\boldsymbol{R}^\prime p^\prime \bar{\sigma}}
 \label{Hint}
 \end{equation} 
where $\hat{N}_{\boldsymbol{R} p  \sigma}=\sum_{\xi} c_{\boldsymbol{R} p \xi \sigma}^{\dagger} c_{\boldsymbol{R} p \xi \sigma} $ is 
the number operator related to the Wannier state. It is also useful to define the \emph{total} number operator $\hat{N}_{\boldsymbol{R}} \!\equiv\! \sum_{p, \sigma} \hat{N}_{\boldsymbol{R} p  \sigma}$. 
Due to spin and orbital degeneracy, the model considers up to 8 electrons per unit cell ($\langle \hat{N}_{\boldsymbol{R}} \rangle \!\leq\! 8$).
As such, one can write the filling factor as $\nu \!=\! \langle \hat{N}_{\boldsymbol{R}} \rangle -4$. 
In passing, we notice that charge neutrality ($\nu=0$) corresponds to $\langle \hat{N}_{\boldsymbol{R}} \rangle\!=\!4$ and that $\nu\!=\!\pm 2$, often referred to in the literature as ``half-filling of the electron (hole) band", corresponds to $\langle \hat{N}_{\boldsymbol{R}} \rangle \!=\!6$  or $\langle \hat{N}_{\boldsymbol{R}} \rangle \!=\!2$ in our calculations.

The Hubbard-like terms in Eq.~\eqref{Hint} encode both short- and long-range density-density interactions. For example, $U^{A A}_{\boldsymbol{R},\boldsymbol{R}} = U^{B B}_{\boldsymbol{R},\boldsymbol{R}} \equiv U$ denotes onsite Hubbard terms, and $U^{A B}_{\boldsymbol{R},\boldsymbol{R}} \equiv U_1$ denotes one of the three nearest-neighbor terms. In the present work, we consider interactions up to 5 nearest neighbors, as depicted in Fig.~\ref{fig:model}(a). 
In addition, for the purposes of this work, we can safely neglect the exchange terms in Eq.~\eqref{Hint}, as they are estimated to be much smaller than the direct ones \cite{koshinoMaximallyLocalizedWannier2018}.

As will become clear in the next section, it is useful to define:
\begin{equation}
    U^{p p^\prime}(\boldsymbol{q}) \equiv \sum_{\boldsymbol{R}, \boldsymbol{R}^\prime  } U^{p p^\prime}_{\boldsymbol{R},\boldsymbol{R}^\prime} e^{i \boldsymbol{\delta}^{p p^\prime}_{\boldsymbol{R},\boldsymbol{R}^\prime} \cdot \boldsymbol{q} } \; ,
\label{eq:Uq}
\end{equation}
where $\boldsymbol{\delta}^{p p^\prime}_{\boldsymbol{R},\boldsymbol{R}^\prime}$ 
are real-space vectors connecting Wannier orbital centers $(p,\boldsymbol{R})$ 
and $(p^\prime,\boldsymbol{R}^\prime)$, see Fig.~\ref{fig:model}(a).

\subsection{Charge and spin fluctuations: matrix-RPA formalism}
\label{sec:RPAformalism}

Different RPA approaches have been used to study the Coulomb screening potential~\cite{pizarroInternalScreeningDielectric2019,goodwinAttractiveElectronelectronInteractions2019,vanhalaConstrainedRandomPhase2020,leconteRelaxationEffectsTwisted2022} and the SC pairing symmetries in TBG~\cite{liuChiralSpinDensity2018,wuHarmonicFingerprintUnconventional2020,huangObservationChiralSlow2022,huangPseudospinParamagnonsSuperconducting2022}. 
The matrix RPA method, for example, can account for pairing vertex diagrams beyond what is usually known as RPA~\cite{altmeyerRoleVertexCorrections2016}.

In particular, several SC gap symmetries are found to be present and to compete as a function of Hubbard and exchange parameters in TBG~\cite{wuHarmonicFingerprintUnconventional2020}, but the chiral $d + id$-wave superconductivity emerges in both RPA~\cite{liuChiralSpinDensity2018,wuHarmonicFingerprintUnconventional2020} and full-scale atomistic modeling with local electronic interactions~\cite{lothmanNematicSuperconductivityMagicangle2022}. 
However, previous works have not considered the moir\'e-scale long-range interactions in MATBG, which have been shown to be large \cite{koshinoMaximallyLocalizedWannier2018,calderonInteractions8orbitalModel2020} and relevant to the superconducting state \cite{huangPseudospinParamagnonsSuperconducting2022}.

Our goal is to probe the SC instability caused by charge and spin quantum fluctuations. 
In the following, we describe the matrix RPA steps used in our analysis. The starting point is the bare (non-interacting) multi-orbital susceptibility matrix elements for each spin \cite{graserNeardegeneracySeveralPairing2009,wuIdentificationSuperconductingPairing2019}: 
\begin{equation}
\begin{split}
    &\left[\hat{\chi}_{0}(\boldsymbol{q},\omega)\right]_{r\xi , t \xi^\prime}^{p \xi, q \xi^\prime} =  \\
    &-\frac{1}{N}\sum_{\boldsymbol{k},\nu\nu'}  \frac{a_{\nu}^{r \xi}(\boldsymbol{k})a_{\nu}^{p \xi}(\boldsymbol{k})^{*}a_{\nu^\prime}^{q \xi^\prime}(\boldsymbol{k}+\boldsymbol{q})a_{\nu^\prime}^{t \xi^\prime}(\boldsymbol{k}+\boldsymbol{q})^{*}}
    {\omega+E_{\nu'}(\boldsymbol{k}+\boldsymbol{q})-E_{\nu}(\boldsymbol{k})+i0^{+}}  \\
    &   \times \left(f(E_{\nu'}(\boldsymbol{k}+\boldsymbol{q}))-f(E_{\nu}(\boldsymbol{k}))\right) \; ,
\end{split}
\label{eq:chi_0td}
\end{equation}
which depend on the eigenvalues $E_\nu (\boldsymbol{k})$ of the non-interacting Hamiltonian 
$H_0$ [Eq.~\eqref{eq:TBmodel}], and on the eigenvector coefficients 
$a_{\nu}^{p \xi}(\boldsymbol{k}) \equiv \langle p \xi | \nu \boldsymbol{k} \rangle$, 
which correspond to the projection of band state $| \nu \boldsymbol{k} \rangle$ into the Wannier orbital $|p \rangle=|A(B) \rangle$  at valley $|{\xi}\rangle=|+(-) \rangle$. As such,  $\hat{\chi}_{0}(\boldsymbol{q},\omega)$ is an $16 \times 16$ matrix spanning  the $\left\{p \xi, q \xi^\prime \right\}$ basis. In Eq. \eqref{eq:chi_0td},
$N$ is the number of BZ $\boldsymbol{k}$-points considered in the summation, and 
$f$ is the Fermi-Dirac distribution for a given temperature $T$. Throughout this work, we used a summation grid of 
98342 $\boldsymbol{k}$-points in the hexagonal lattice BZ.

Following Refs.~\cite{esirgenMathitWavePairing1999a,graserNeardegeneracySeveralPairing2009,luoNeutronARPESConstraints2010,nicholsonRoleDegeneracyHybridization2011,nicholsonCompetingPairingSymmetries2011,martinsRPAAnalysisTwoorbital2013a,wuIdentificationSuperconductingPairing2019}, we write the RPA spin and charge susceptibilities suitable to probe for magnetism 
and/or charge order in the system, respectively, by 
\begin{align}           
    \hat{\chi}_{\text{s}}(\boldsymbol{q},\omega)	=\hat{\chi}_{0}(\boldsymbol{q},\omega)\left[\hat{1}-\hat{U}_{s}(\boldsymbol{q})\hat{\chi}_{0}(\boldsymbol{q},\omega)\right]^{-1}, \label{eq:chi_s} \\
    \hat{\chi}_{\text{c}}(\boldsymbol{q},\omega)	=\hat{\chi}_{0}(\boldsymbol{q},\omega)\left[\hat{1}+\hat{U}_{c}(\boldsymbol{q})\hat{\chi}_{0}(\boldsymbol{q},\omega)\right]^{-1}, \label{eq:chi_c}
\end{align} 
where the non-zero $\hat{U}_{s}(\boldsymbol{q})$ and $\hat{U}_{c}(\boldsymbol{q})$ matrix elements in terms of the $U^{p p^\prime}(\boldsymbol{q})$ defined in Eq.~\eqref{eq:Uq} are given by \footnote{A more detailed derivation of the interaction matrices is presented in Ref.~\cite{brazChargeSpinFluctuations2024}.}: 
\begin{align}
\left[ \hat{U}_{c}(\boldsymbol{q})\right]^{p \xi , p\xi' }_{p \xi , p\xi' } &= U^{p p}(\boldsymbol{q}) \delta_{\xi \xi'}  \; , \label{eq:Uc} \\
\left[ \hat{U}_{c}(\boldsymbol{q})\right]^{p \xi, p \xi'}_{r \xi, r \xi' } &= 2U^{pr}(\boldsymbol{q})  \delta_{\xi \xi'}  \; , \label{eq:Ucpr}\\
\left[ \hat{U}_{s}(\boldsymbol{q})\right]^{p \xi , p\xi' }_{p \xi , p\xi' } &= U^{pp}(\boldsymbol{q})  \delta_{\xi \xi'} 
\label{eq:Us} \; ,
\end{align}
where $p \neq r$ in Eq. \eqref{eq:Ucpr}.

In this work, we consider long-range interactions up to the fifth 
nearest neighbors, making use of the next-neighbors ratios 
calculated in Table I of Ref.~\cite{koshinoMaximallyLocalizedWannier2018} namely,  
$U_i/U_1=0.7469, 0.6967, 0.4547, 0.4005~~(i=2,\ldots,5)$, 
while $U_1/U$ and $U$ will be taken as free parameters \footnote{We note that our main results hold over a range of values for the $U_i/U$ ratios. For instance, while the ratios of Ref.~\cite{koshinoMaximallyLocalizedWannier2018} result in $U_5/U_2=0.5362$, we find that the results of Figs.~\ref{fig:lambda_filling} and \ref{fig:RPA_chi} will hold even for $U_5/U_2 \sim 0.35$. See Appendix \ref{appendix:ratios} for a full discussion. }.

Due to the strong Hubbard-like interactions, a spin-singlet pairing mechanism 
is believed to be the leading candidate for superconductivity in MATBG \cite{fischerSpinfluctuationinducedPairingTwisted2021, lothmanNematicSuperconductivityMagicangle2022},  
as there is currently no hard experimental evidence for spin-polarized Copper pairs \cite{caoUnconventionalSuperconductivityMagicangle2018,wuIdentificationSuperconductingPairing2019} and some theoretical studies 
suggest that singlet spin fluctuations can lead to SC pairing in MATBG \cite{fischerSpinfluctuationinducedPairingTwisted2021}. 
In addition, previous RPA calculations show that spin-singlet superconductivity is more prominent in MATBG~\cite{zhangDensityWaveTopological2020},
while, more recently, the onset of SC phases originating from interactions between electrons on the same honeycomb sublattice has been found to be of the spin-singlet type~\cite{huangPseudospinParamagnonsSuperconducting2022}.

We proceed with the calculation of the spin-singlet multiorbital pairing 
vertex 
$\left[\Gamma(\boldsymbol{k},\boldsymbol{k}',\omega) \right]_{r\xi, t\xi^\prime}^{p\xi, q\xi^\prime}$
related to the RPA charge and spin susceptibilities as \cite{esirgenMathitWavePairing1999a,graserNeardegeneracySeveralPairing2009,wuIdentificationSuperconductingPairing2019}:
\begin{align}
\begin{split}
        \left[\Gamma(\boldsymbol{k},\boldsymbol{k}',\omega) \right]_{r\xi , t \xi^\prime}^{p \xi, q \xi^\prime}  & =\Bigg{[}\frac{3}{2}\hat{U}_{s}\hat{\chi}_{s}(\boldsymbol{k}-\boldsymbol{k}',\omega)\hat{U}_{s}+\frac{1}{2}\hat{U}_{s} \\
        &  -\frac{1}{2}\hat{U}_{c}\hat{\chi}_{c}(\boldsymbol{k}-\boldsymbol{k}',\omega)\hat{U}_{c}+\frac{1}{2}\hat{U}_{c} \Bigg{]}_{p \xi, r\xi }^{ t \xi^\prime, q \xi^\prime} 
 \; , \label{vertex}
\end{split}
\end{align} 
where $\hat{\chi}_{s}(\boldsymbol{k})$ and $\hat{\chi}_{c}(\boldsymbol{k})$ are the spin and charge susceptibilities defined in Eqs.~\eqref{eq:chi_s} and \eqref{eq:chi_c}, respectively, while $p,q,r,t$ are orbital indices [as in Eq.~\eqref{eq:chi_0td}].

We also define the kernel function $\Gamma(\boldsymbol{k},\boldsymbol{k}')$, associated to the scattering of a singlet pair $(\boldsymbol{k} \uparrow p, -\boldsymbol{k} \downarrow r)$ at a FS sheet  $C_i$ 
with another pair $(\boldsymbol{k}^\prime \uparrow q, -\boldsymbol{k}' \downarrow t)$ at another (disconnected) FS sheet $C_{j}$ \cite{wuIdentificationSuperconductingPairing2019}: 
\begin{align}
\begin{split}
        \Gamma(\boldsymbol{k},\boldsymbol{k}')=& \sum_{rtpq, \xi \xi^\prime}a_{\nu_{-k}}^{t \xi^\prime,*}(-\boldsymbol{k})a_{\nu_k}^{r\xi,*}(\boldsymbol{k})   \\
        &  \times \text{Re}[\Gamma_{r\xi , t \xi^\prime}^{p \xi, q \xi^\prime}(\boldsymbol{k},\boldsymbol{k}',0)]a_{\nu_{k'}}^{p \xi}(\boldsymbol{k}')a_{\nu_{-k'}}^{q \xi^\prime}(-\boldsymbol{k}') \; .        
\end{split} 
\label{vertex_full}
\end{align}
where $a_{\nu_k}^{p \xi}$ are defined as in Eq.~\eqref{eq:chi_0td} with the caveat that the band index $\nu_k$ is now defined by the FS constraint $E_{\nu_k}(\boldsymbol{k}) = E_F$, where $E_F$ is the Fermi energy. 

The kernel $\Gamma(\boldsymbol{k},\boldsymbol{k}')$ [Eq.~\eqref{vertex_full}] is a matrix having $N_k$ rows and $N_{k'}$ columns so there will be $N_k$ eigenvalues, one for each $\boldsymbol{k}$.
Thus, $\Gamma(\boldsymbol{k},\boldsymbol{k}')$ is associated with the pairing strength $\lambda_{\alpha}$, which relates to 
the gap function $g_{\alpha}(\boldsymbol{k})$ through the integral equation~\cite{takimotoStrongcouplingTheorySuperconductivity2004,graserNeardegeneracySeveralPairing2009}
\begin{equation}
    -\sum_{j}\oint_{C_{j}}\frac{d\boldsymbol{k}'_{||}}{v_{F}(\boldsymbol{k}')}\frac{1}{(2\pi)^{2}}\bar{\Gamma}(\boldsymbol{k},\boldsymbol{k}')g_{\alpha}(\boldsymbol{k}')=\lambda_{\alpha} g_{\alpha}(\boldsymbol{k}) \; ,
\label{gap}
\end{equation} 
where the $\bar{\Gamma}(\boldsymbol{k},\boldsymbol{k}')$ 
is the so-called symmetric part of the $\Gamma(\boldsymbol{k},\boldsymbol{k}')$ 
kernel function defined in Eq.~\eqref{vertex_full}: 
$\bar{\Gamma}(\boldsymbol{k},\boldsymbol{k}') \equiv [\Gamma(\boldsymbol{k},-\boldsymbol{k}')+\Gamma(\boldsymbol{k},\boldsymbol{k}')]/2$  (see Appendix~\ref{appendix:pairing_strength} for a formal derivation). 

The summation and integral operators in Eq.~\eqref{gap} have the same role as the summation in a usual eigenvalue equation. 
The eigenvalues $\lambda_{\alpha}$ and 
eigenvectors $g_{\alpha}(\boldsymbol{k})$ refer to the momentum vectors 
$\boldsymbol{k}$ probed over the respective FS 
$C_{i}$,
to which the line/surface integral refers. 
In other words, the number of points probing the FS of the system is, therefore, the dimension of the matrix 
$\frac{d\boldsymbol{k}'_{||}}{v_{F}(\boldsymbol{k}')}\frac{1}{(2\pi)^{2}}\Gamma(\boldsymbol{k},\boldsymbol{k}')$ that shall be diagonalized. 
The largest superconducting critical temperature ($T_c$) value 
corresponds to the leading pairing strength $\lambda \equiv \max ( \lambda_{\alpha} )$, 
which will dominate the SC instability.

The generalized Stoner criterion  (namely, the vanishing of the denominator in the RPA spin and charge susceptibilities at $\omega=0$, Eqs.~\eqref{eq:chi_s} and \eqref{eq:chi_c} respectively) establishes the condition for the transition between a paramagnetic (uniform density) state possibly favoring the SC phase, and a magnetically (charge) ordered one in the particle-hole (particle-particle) channel
\cite{berkEffectFerromagneticSpin1966a,bickersConservingApproximationsStrongly1989b,bickersConservingApproximationsStrongly1991,takimotoStrongcouplingTheorySuperconductivity2004,sakakibaraOriginMaterialDependence2012}. 
As such, within the RPA approach, the Stoner criterion can be used as proxy to identify the onset of a magnetic-ordered phase in the particle-hole channel \cite{Engstroem:Phys.Rev.B:014508:2023}. With this goal, we
define the spin ($\alpha_s$) and charge ($\alpha_c$) Stoner parameters by solving the 
following eigenvalue equations~\cite{sakakibaraOriginMaterialDependence2012} 

\begin{equation}
\begin{split}
    & \hat{1}\alpha_s - \hat{U}_s\hat{\chi}_0 = 0, \\
    & \hat{1}\alpha_c + \hat{U}_c\hat{\chi}_0 = 0.
\end{split}
\label{stoner}
\end{equation}
The matrix-RPA Stoner criterion is fulfilled when $\max \{ \alpha_s,\alpha_c \} =1$.

\section{Superconducting pairing vertex}
\label{sec:pairingvertex}

%-----------F I G U R E  2  ------
\begin{figure}[t]
\begin{center}
\includegraphics[width=1.0\columnwidth]{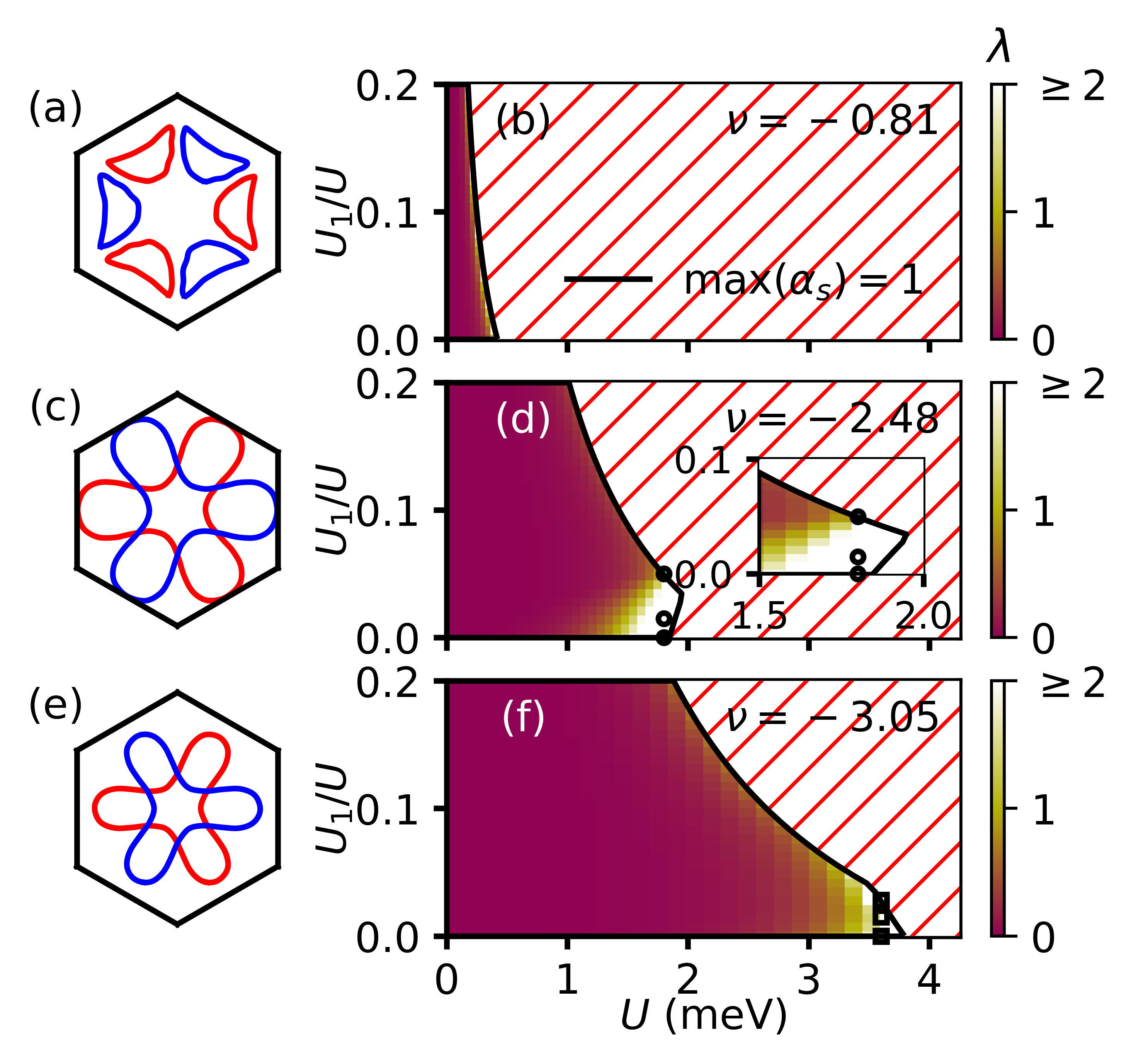}
	\caption{Panels (a), (c), and (e) show Fermi surface plots for the $\xi=+$ 
	(blue) and $\xi=-$ (red) orbitals, for different fillings $\nu$, 
	as indicated in Fig.~\ref{fig:model}(b) by 
	horizontal dashed lines. Panels (b), (d), and (f): phase diagram color maps denoting 
	the pairing strength $\lambda$ as a function of $U_1/U$ and $U$, for $T=1$ mK. 
	The 
 %red dashed 
        black line in each panel indicates where the magnetic Stoner criterion 
	has been fulfilled, i.e., $\mbox{max}(\alpha_s)=1$, resulting in magnetic order 
	to its right side (hatched area). 
 %We also show the $\lambda=1$ (strong coupling) curve for reference.
	We clipped values above $\lambda=2$ to white color. 
	The symbols in panels (d) and (f) denote the interaction 
	values for which the pairing vertices shown in Fig.~\ref{fig:RPA_chi} 
	were computed.
}
\label{fig:lambda_filling}
\end{center}
\end{figure}
%-----------E N D  F I G U R E   ------

We now turn to the 
connection between short- and long-range interactions and the onset of superconductivity in MATBG. 
Concretely, we will consider the hole-doped case, with Fermi energies [$E_F$s, depicted by the horizontal lines in Fig.~\ref{fig:model}(b)] located at or below the lower (hole-like) Van Hove singularity shown in  Fig.~\ref{fig:model}(c).

The Fermi surfaces for three different filling factors 
are shown in Figs.~\ref{fig:lambda_filling}(a), (c), and (e).
In Fig.~\ref{fig:lambda_filling}(a), we show the (hole) pockets 
holding scattering events for $\nu=-0.81$, which corresponds 
to an $E_F$ value pinned at the Van Hove singularity. 
Panels (c) and (e) in Fig.~\ref{fig:lambda_filling} present 
FS energy cuts for $\nu=-2.48$ and $\nu=-3.05$, which have lobes contained 
within the mini-BZ. In fact, for $\nu=-2.48$ the lobes of the 
FS are touching the $M$ points at the edges of the BZ.

We solve Eq.~\eqref{gap}, for the FSs shown in Figs.~\ref{fig:lambda_filling}(a), 
(c), and (e), as a function of the Hubbard parameter $U$ and the long-range control 
parameter $U_1/U$.
The results are shown in panels (b), (d), and (f) in Fig.~\ref{fig:lambda_filling}.

The previously defined leading pairing strength $\lambda=\max ( \lambda_{\alpha} )$ quantifies 
the highest SC critical temperature ($T_c \propto e^{-1/\lambda}$). 
Thus, in Fig.~\ref{fig:lambda_filling}, $\lambda \lesssim 1$ regions (purple to yellow colors) indicate SC phases with moderate-to-high $T_c$, while $\lambda \geq 1$ (yellow to white colors) means strong coupling and potentially larger $T_c$ values. 
%We show the $\lambda=1$ curve for reference (black solid curve). 
Magnetically or charge-ordered states are obtained when the spin or charge fluctuations generate a condensate with a critical temperature higher than $T_c$.
This happens when the maximum Stoner parameter (either $\alpha_s$ or $\alpha_c$) [Eqs.~\eqref{stoner}] becomes equal to $1$. 
As indicated in Fig.~\ref{fig:lambda_filling}, within the range of the parameters explored here, spin 
order 
%(red dashed curve) 
(black curve) always emerges before charge order (see Appendix \ref{appendix:stoner} for details).

%-----------F I G U R E  RPA chi ------
\begin{figure}[t]
\begin{center}
\includegraphics[width=1.0\columnwidth]{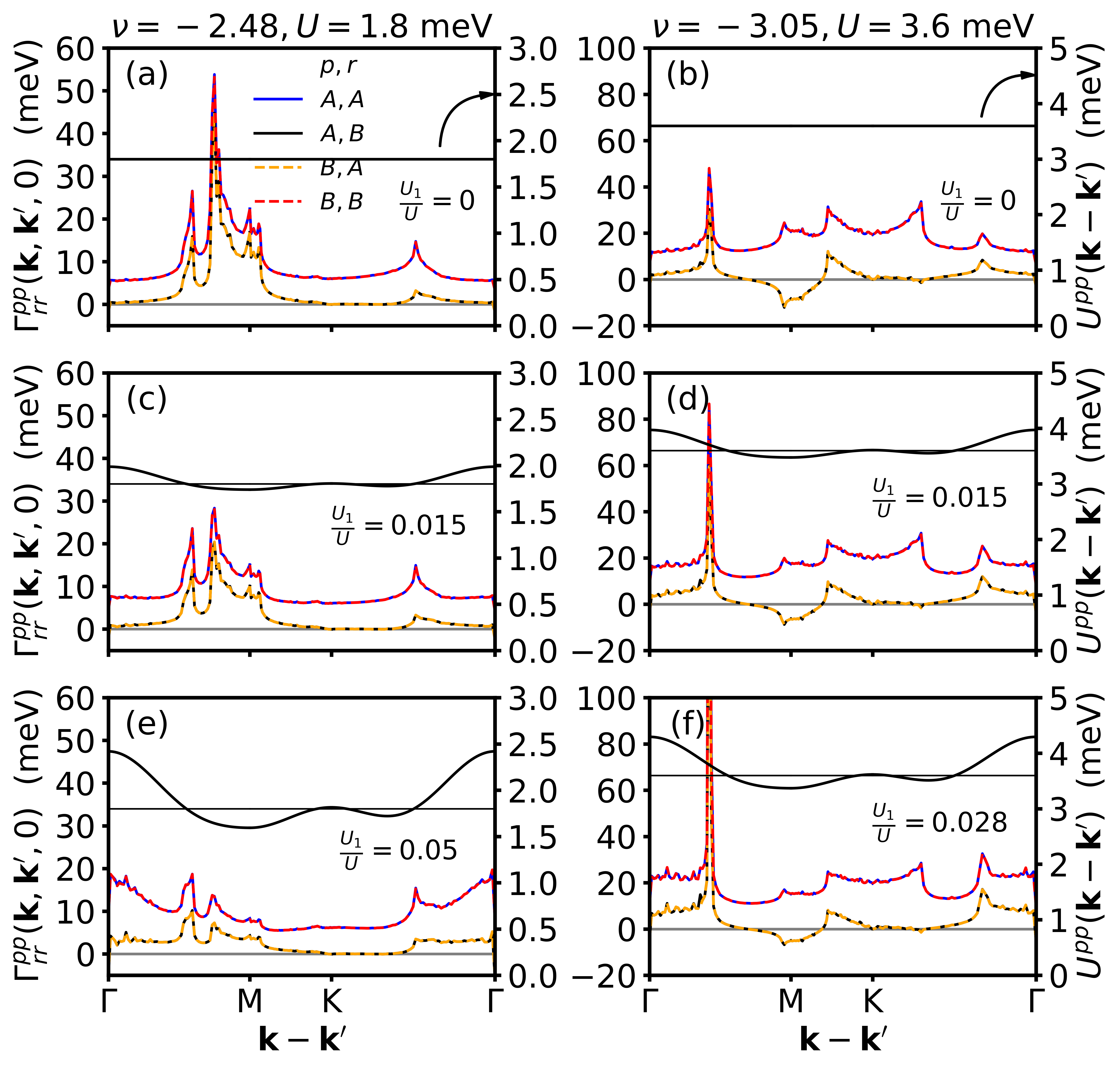}
\caption{
	Spin-singlet pairing vertex [Eq.~\eqref{vertex}] (left axis) matrix elements and momentum-dependent Hubbard-like terms [Eq.~\eqref{eq:Uq}] (right axis) for 
	$\nu=-2.48$ at $U=1.8$ meV and $\nu=-3.05$ at $U=3.2$ meV (left and right panels, respectively), 
	and $U_1/U=0$ (a,b), $U_1/U=0.015$ (c,d), or $U_1/U=0.055$ (e,f).
	The left hand panels correspond to the circle symbols and the right hand panels to square symbols in the phase diagrams shown in Fig.~\ref{fig:lambda_filling}(d) and (f), respectively.
 The thin horizontal black lines 
 denote the Hubbard $U$ [Eq.~\eqref{eq:Uq} at $U_1/U=0$] in each simulation. 
 }
\label{fig:RPA_chi}
\end{center}
\end{figure}
%-----------E N D  F I G U R E RPA_chi  ------

%\LUIS{Did some rewriting in light of the new Fig. 2}

% The relative position of the $\lambda=1$ (black solid) and 
% $\max(\alpha_s)=1$ (red dashed) curves in Fig.~\ref{fig:lambda_filling} 
% varies with $\nu$. Indeed, they are practically superposed for $\nu=-0.81$ 
% filling [panel (b)], while they are slightly displaced in relation to 
% one another for the two other fillings [panels (d) and (f)]. 

The $\max(\alpha_s)=1$ displays a ``duckbill shape" 
for fillings close to $\nu=-2.48$ (Fig.~\ref{fig:lambda_filling}-(d))
%, where the curve for $\lambda=1$ 
%(as a function of $U_1/U$) has a maximum at finite $U_1/U$ [the same occurs 
%for the $\max(\alpha_s)=1$ curve in panel (d)]. This has important implications, 
%as described next. 
The presence of the ``duckbill shape" 
%(for both the solid and the  dashed lines) 
in Fig.~\ref{fig:lambda_filling}(d) generates a particular 
region of the phase diagram that may be described as follows: (i) moving from left to right along a horizontal line passing through either of the two circles at $U_1/U=0$ and $0.015$ will result in an exponential increase in $T_c$ 
with $U$, creating a region in the phase diagram (around the ``duckbill'' shape) with the most robust 
superconductivity, \emph{before} magnetic order sets in 
(this kind of behavior will occur as long as $U_1/U \lesssim 0.04$); (ii) a 
vertical line passing through the circles  will result in an 
increase in $T_c$, for decreasing $U_1/U$.
This last behavior, in particular, 
will be analyzed in more detail below 
(see Fig.~\ref{fig:RPA_chi}), through a study of the most relevant 
spin-singlet pairing vertex matrix elements for $U$ and $U_1/U$ values marked by the symbols in 
Fig.~\ref{fig:lambda_filling}(d) (circles) and (f) (squares).

Figure~\ref{fig:RPA_chi} shows the intersublattice 
and intrasublattice spin-singlet pairing vertex matrix 
elements [Eq.~\eqref{vertex}] computed for $\nu=-2.48$. 
Panels in the left (right) column present results for the $U_1/U$ and $U$ values corresponding to the symbols marked in 
Fig.~\ref{fig:lambda_filling}(d) [(f)]. 
We first note that curves corresponding to intersublattice matrix elements ($A,B$ and $B,A$) are almost identical, and the same is true for the intrasublattice ones 
($A,A$ and $B,B$). 

Moreover, the values of  the intrasublattice matrix elements are systematically 
larger than the intersublattice ones. For most scattering momenta, they are both positive (repulsive effective interaction), although 
a small negative (attractive effective interaction) region appears for  $\nu=-3.05$ filling [panels (b) and (d)],
around the $\boldsymbol{M}$ high-symmetry point for the 
intersublattice matrix elements.
Based in these results,  we argue that the main SC channel in the strong-coupling region is of the singlet type, 
since our pairing vertex calculations for the triplet channel are lower 
in absolute value than for the singlet channel, although they 
display larger negative value (attractive) regions
(see Appendix~\ref{appendix:triplet}).  

In addition to the pairing vertex, Fig.~\ref{fig:RPA_chi} also presents results for the 
intrasublattice interactions $U^{pp}(\boldsymbol{k}-\boldsymbol{k}')$ (thick black lines, with its value 
displayed in the right-side vertical axis). These intrasublattice interactions are the most relevant to our analysis because the spin fluctuations, which dominate pairing, depend only on them [Eq.~\eqref{eq:Us}]. We note that, 
for $U_1/U=0$, $U^{pp}$ does not depend on $\boldsymbol{k}-\boldsymbol{k}'$ [panels (a) and (b)]. Those horizontal lines 
are repeated on the finite $U_1/U$ panels to highlight the 
fact that, once long-range interactions are turned on, the BZ 
divides itself up into regions where repulsion is increased 
(a large region around the $\Gamma$ point and small regions 
around the $K$ points) or decreased 
(the rest of the BZ, see Fig.~\ref{fig:nestings}) in relation to the constant repulsion value for vanishing $U_1/U$. In what follows, we will refer to the latter region as having `attractive' long-range components and the former as having `repulsive' long-range components. As discussed 
below, the existence of the duckbill shape for an specific 
filling depends on which region (attractive or repulsive) 
is located the nesting vector, for $U_1/U=0$, where the 
divergence of both the spin susceptibility and the pairing 
vertex occur.

\section{ Discussion on the ``duckbill'' feature in the Stoner boundary line. }
\label{sec:duckbill}

\subsection{The role of nesting }
\label{sec:nesting}

The results shown in the left-side panels of Fig.~\ref{fig:RPA_chi}, for $\nu=-2.48$, indicate that the pronounced peak in the pairing vertex matrix elements $\Gamma^{p\xi p\xi'}_{r\xi r\xi'}(\boldsymbol{q})$ around the $M$ point ($\boldsymbol{q} \approx \boldsymbol{q}_{\boldsymbol{M}}$) are suppressed as $U_1/U$ increases. 
Here, we argue that, for small values of $U_1/U$, such suppression is related to the interplay of strong nesting effects at $\boldsymbol{q} \approx \boldsymbol{q}_{\boldsymbol{M}}$ and the fact that there are effectively negative (attractive) interaction contributions that  make diagonal interaction terms $U^{pp}(\boldsymbol{q})$  [defined by Eq.~\eqref{eq:Uq}] to be smaller than $U$, precisely around $\boldsymbol{q}=\boldsymbol{q}_{\boldsymbol{M}}$, due to the lattice geometry.

\begin{widetext}

%%%%%%%%%%%%%%%%%%%%%%%%%%%%%%%%%%%%%%%%%%%%%%%%%%%%%%%%%%%%%%%%%%%%%%%%%%%%%%%%
\begin{figure*}[ht]
\includegraphics[width=0.98\columnwidth]{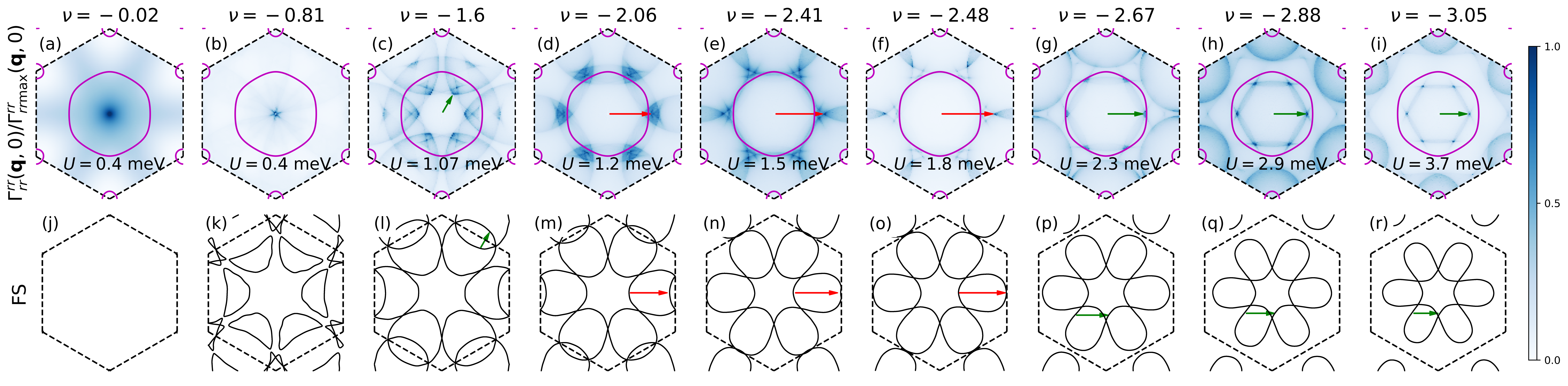}
\caption{ 
    Spin-singlet pairing vertex [Eq.~\eqref{vertex}] (top row) for $U_1=0$ and different values of $U$ and respective non-interacting Fermi surfaces (lower row) for different fillings $\nu$.
    The curves defined by $U^{pp}(\boldsymbol{q})=U$ 
    are shown in magenta, with $U^{pp}(\boldsymbol{q})<U$ in the inner part of the pockets. 
    The arrows denote the nesting vectors and which regions of the Fermi surfaces they connect.
    Red arrows highlight fillings with nesting vectors near the $M$ point.
}
\label{fig:nestings}
\end{figure*}
%%%%%%%%%%%%%%%%%%%%%%%%%%%%%%%%%%%%%%%%%%%%%%%%%%%%%%%%%%%%%%%%%%%%%%%%%%%%%%%%

\end{widetext}

Figure \ref{fig:nestings} shows the normalized diagonal pairing vertex $\Gamma_{rr}^{rr}(\boldsymbol{q},\omega=0)$  (see Eq.~\eqref{vertex}; for clarity, we omit the valley indices in the following notation) for different filling factors (upper panels), along with the corresponding Fermi surfaces (lower panels). As a general trend, the peaks in $\Gamma_{rr}^{rr}(\boldsymbol{q},\omega=0)$ occur at the FS nesting vectors shown in the lower panels.  Notice that, for $\nu \approx -2.48$, the nesting is close to the $M$ point and, for finite 
$U_1/U$, located inside the $U^{pp}(\boldsymbol{q})<U$ region, meaning that the overall contribution of the long-range interaction terms to $U^{pp}(\boldsymbol{q})$ is \emph{negative} (attractive). 

From these results, the qualitative behavior of the Stoner boundary [black line in Fig.~\ref{fig:lambda_filling}(d)] can be understood as follows. Let us consider the leading diagonal terms of the non-interacting susceptibility [$\left[\hat{\chi}_0(\boldsymbol{q})\right]_{pp}^{pp}$  from Eq.~\eqref{eq:chi_0td}] and the interaction matrix $U^{pp}_s(\boldsymbol{q})$ [Eq.~\eqref{eq:Uq}]. For $U_1/U\approx0$ the Stoner criterion [$\mbox{max}(\alpha_s)=1$ in Eq.~\eqref{stoner}] is dominated by  $\left[\hat{\chi}_0(\boldsymbol{q} \approx \boldsymbol{q}_{\boldsymbol{M}})\right]_{pp}^{pp} \sim 1/U$, where the maximum in $\left[\hat{\chi}_0(\boldsymbol{q})\right]_{pp}^{pp}$ occurs due to nesting at $\boldsymbol{q}\approx\boldsymbol{q}_{\boldsymbol{M}}$, since, for small $U_1/U$,  $U^{pp}_s(\boldsymbol{q}) \approx U$ (and largely independent of $\boldsymbol{q}$). This analysis applies to the immediate 
vicinity of the $U_1/U=0$ circle in Fig.~\ref{fig:RPA_chi}(d), 
which sits very close to the Stoner boundary.

As one moves away from the Stoner boundary into the superconducting phase by \emph{increasing} $U_1/U$, while keeping $U$ fixed, $U^{pp}_s(\boldsymbol{q})$, at $\boldsymbol{q} \approx \boldsymbol{q}_{\boldsymbol{M}}$, will actually be \emph{reduced}. Thus, in order to ``return'' to the Stoner boundary while keeping $U_1/U$ fixed, one has to \emph{increase} $U$ in order to fulfill the Stoner criterion again. Thus, the Stoner boundary, for small values of $U_1/U$, will have a \emph{positive} slope in the $U_1/U$ vs. $U$ plane.

Notice that the main condition for such behavior to occur is precisely that $U^{pp}_s(\bar{\boldsymbol{q}}) < U$ at the nesting vector $\bar{\boldsymbol{q}}$, a situation occurring for fillings near $\nu=-2.48$ (precisely where the lobes of the FS touche the $M$ point of the BZ) with $\bar{\boldsymbol{q}} \approx \boldsymbol{q}_{\boldsymbol{M}}$  [see Figs.~\ref{fig:nestings}(d-f)]. 
These nesting vectors are indicated by red arrows in 
Fig.~\ref{fig:nestings}.
For other filling factors, the nesting occurs at regions where, for 
finite $U_1/U$, $U^{pp}_s(\bar{\boldsymbol{q}}) > U$ (``inside'' the closed magenta curves defined by $U^{pp}_s(\boldsymbol{q}) = U$, as shown in the upper panels of Fig.~\ref{fig:nestings}), which produces a negative slope for the Stoner boundary  in the $U_1/U$ vs. $U$ plane. 
The corresponding nesting vectors are indicated by green 
arrows in Fig.~\ref{fig:nestings}. We emphasize that the physical picture presented here is quite generic and depends only on the interplay of nesting effects and the structure of $U^{pp}(\boldsymbol{q})$, which is largely model-independent.

Now, for \emph{large} values of $U_1/U$, $U^{pp}_s(\boldsymbol{q})$ has a strong $\boldsymbol{q}$-dependence and a pronounced maximum at $\boldsymbol{q}=0$ [see, e.g.,  Fig.~\ref{fig:RPA_chi}(e)]. Thus, the Stoner criterion can be fulfilled by $\left[\hat{\chi}_0(\boldsymbol{0})\right]_{pp}^{pp} \sim 1/U^{pp}_s(\boldsymbol{0})$, independently of nesting conditions. Since the overall contributions of the long-range interactions to $U^{pp}_s(\boldsymbol{q})$ are positive (repulsive) at $\boldsymbol{q}\approx \boldsymbol{0}$, this implies that as one moves away from the Stoner boundary and into the SC region by \emph{decreasing} $U_1/U$, this will also \emph{reduce} $U^{pp}_s(\boldsymbol{q}\approx \boldsymbol{0})$.  Again, one would have to  \emph{increase} $U$, while keeping this value of $U_1/U$ fixed, in order to re-establish the conditions for the Stoner criterion. Thus, the Stoner boundary will generally have a \emph{negative} slope in the $U_1/U$ vs. $U$ plane for large values of $U_1/U$, as shown in Figs.~\ref{fig:RPA_chi}(b,d,f).

% ------------------------------
\subsection{Stoner parameters }
\label{appendix:stoner}

\begin{figure}[t]
\begin{center}
\includegraphics[width=1.0\columnwidth]{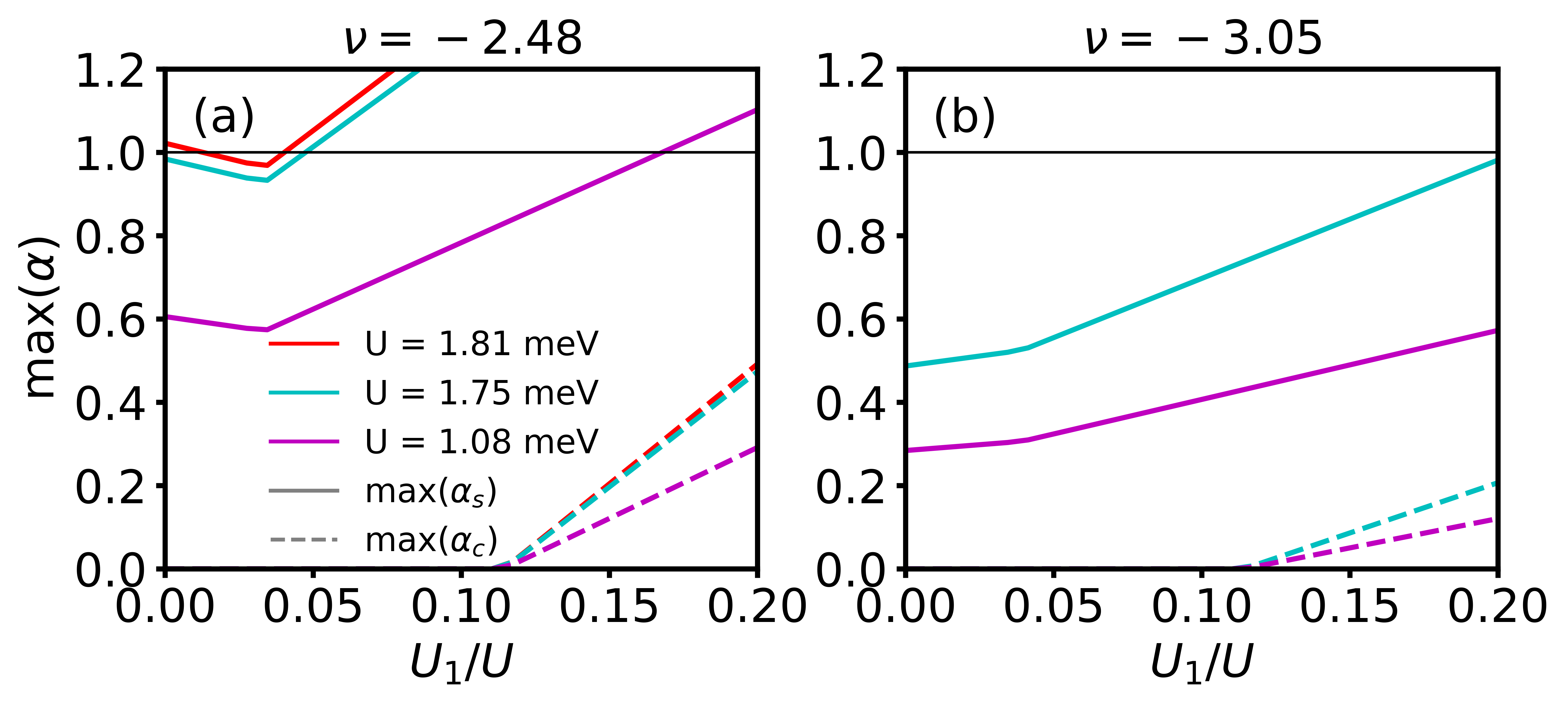}
\caption{
	Stoner parameters
 $\alpha_s$ ($\alpha_c$), for fillings 
	$\nu=-2.48$ (a) and $\nu=-3.05$ (b),
and different values of $U$ as a function  of long-range coupling strength $U_1/U$. 
}
\label{fig:stoner}
\end{center}
\end{figure}

The above arguments can be put into more quantitative terms by plotting the Stoner criterion parameters $\alpha_s$ and $\alpha_c$ defined in Eq.~\eqref{stoner} as a function of $U_1/U$ for given values of $U$ and $\nu$. Figure~\ref{fig:stoner} shows the spin 
($\mbox{max}\left(\alpha_s \right)$, solid lines) and charge ($\mbox{max}\left( \alpha_c \right)$, dashed lines) 
Stoner parameters [Eq.~\eqref{stoner}] computed as a function 
of $U_1/U$, for fillings $\nu=-2.48$ [panel (a)] 
and $\nu=-3.05$ [panel (b)] and different values of $U$.

For $\nu=-2.48$ [Fig.~\ref{fig:stoner}(a)], $\mbox{max}\left( \alpha_s \right)$ shows a non-monotonic behavior as a function of $U_1/U$ for all values of $U$ shown. It initially \emph{decreases} with $U_1/U$, reaches a minimum around $U_1/U\! \approx \!0.04$, and increases again, until the Stoner instability ($\mbox{max}\left(\alpha_s\right)\! =\! 1$) is reached. 
By contrast, for $\nu=-3.05$, $\mbox{max}\left(\alpha_s\right)$ shows a monotonic behavior as a function of $U_1/U$, although with a ``cusp-like'' feature around $U_1/U\! \approx \!0.04$.

%-----------F I G U R E  5 ------
%-----------E N D  F I G U R E  ------

More interestingly, for $U=1.81$ meV, $\mbox{max}\left(\alpha_s\right)$ lies \emph{above} the Stoner instability value already at $U_1/U \!=\! 0$.  Surprisingly, \emph{increasing} $U_1/U$ from zero tends to suppress the magnetic order and favor the SC phase at $U_1/U\! \approx \!0.01$, crossing the Stoner instability line \emph{from above}. As expected, by further increasing $U_1/U$ one crosses the Stoner instability line from below at $U_1/U\! \approx \!0.04$.  This is a signature of the ``duckbill'' feature discussed earlier.

We notice in passing that the charge Stoner parameter $\mbox{max}\left(\alpha_c\right)$ (dashed lines in Fig.~\ref{fig:stoner})
is essentially negligible in the
$0 \leq U_1/U \leq 0.2$ interval, for both cuts and both fillings.

\section{Concluding remarks}
\label{sec:Conclusions}

In this work, we investigated the interplay between on-site and long-range interactions in the SC phase of MATBG. By employing a low-energy model which accurately describes the low-lying flat bands at the magic angle and by performing matrix-RPA calculations, we have uncovered intriguing insights into the crucial role played by long-range interactions in shaping the SC behavior of this material.

Our investigations reveal a novel and intricate dependence of the SC phase on the strength of long-range interactions. We observe that, as a general trend, both on-site and long-range interactions can favor the emergence of the SC phase up to a certain threshold when a Stoner-like instability sets in.  
Notably, for some band-filling factors where nesting between momentum points at distances close enough to the $M$ points in the Brillouin zone is present in the Fermi surface, our results showcase a distinctive feature in the phase diagram -- a ``duckbill shape'' in the Stoner boundary.

Such surprising ``reentrant behavior" in the phase diagram offers a scenario in which \emph{increasing} the strength of long-range interactions relative to the local ones can drive the system across the Stoner boundary and into the SC phase and underscores the intricate sensitivity of the SC state to the delicate balance between short- and long-range interactions.
We emphasize that this feature is the result of an interplay of Fermi surface nesting and lattice geometry effects on the long-range interactions (which include attractive components around the $M$ points) and it constitutes a rather generic result which can be extended to systems featuring similar lattice geometries.

In summary, this work underscores the role of the non-trivial interplay between magnetism and long-range interactions in the unconventional SC behavior of MATBG. We leave for future work a more detailed investigation of the SC gap symmetry as a function of filling and long-range interaction strength, and the possible magnetic textures in the phase diagram. 

We have made this model openly available in Python and Fortran \cite{brazCodeMatrixRPA2024}.

\begin{acknowledgments}
L.G.G.V.D.S. acknowledges financial support by CNPq (423137/2018-2, and 309789/2020-6), and FAPESP (grant No. 2016/18495-4). L.B.B. acknowledges financial support by the Coordena\c{c}\~ao de Aperfei\c{c}oamento de Pessoal de N\'{\i}vel Superior - Brasil (CAPES) - Finance Code 001.
\end{acknowledgments}

% Appendix -------------------------------------------
\appendix
% ------------------------------
\section{Derivation of the pairing strength equation}
\label{appendix:pairing_strength}

Here, we provide a derivation of Eq.~\eqref{gap}. First, equations \eqref{eq:Uc}-\eqref{eq:Us} are obtained by explicitly identifying the terms of Eq.~\eqref{Hint} to those of a generic (multisublattice) interaction Hamiltonian, whose terms can be tracked down as contributions to the RPA spin and/or charge susceptibilities (Eqs.~\eqref{eq:chi_s} and \eqref{eq:chi_c}). These, in turn, are used to compute the expression for the pairing vertex $\Gamma^{pq}_{rt}(\boldsymbol{k},\boldsymbol{k}',\omega)$ (Eq.~\eqref{vertex}), which is sensible to the (non-interacting) electronic structure of the system.

Next, we write the effective pairing potential, which encodes the dynamical screening of the system's interacting potential \cite{wuHarmonicFingerprintUnconventional2020}, given by:
\begin{equation}
    V_{\text{eff}}(\omega)=\sum_{\substack{\boldsymbol{k}\boldsymbol{k}' \\ pqrt \\ \xi \xi^\prime}  }\Gamma_{r\xi , t \xi^\prime}^{p \xi, q \xi^\prime}(\boldsymbol{k},\boldsymbol{k}',\omega)c_{\boldsymbol{k}p \xi \uparrow}^{\dagger}c_{-\boldsymbol{k}r \xi\downarrow}^{\dagger}c_{-\boldsymbol{k}'t \xi^\prime \downarrow}c_{\boldsymbol{k}'q \xi^\prime\uparrow}  \; .    
\label{screening}
\end{equation}

Following Scalapino \textit{et al.} \cite{scalapinoFermisurfaceInstabilitiesSuperconducting1987a}, we define Fermi-surface averaged spectral weights
\begin{equation}
	    F(\omega)=-\frac{1}{\pi}\sum_{\boldsymbol{k}'}\oint_{C_{j}}\frac{d\boldsymbol{k}'}{(2\pi)^{d}v_{F}(\boldsymbol{k}')}\text{Im}\bar{\Gamma}(\boldsymbol{k},\boldsymbol{k}',\omega),
\label{spectral_weights}
\end{equation}
where $d$ is the momentum space dimension and the integral is taken over closed Fermi surfaces $C_{j}$ present on the Fermi level of the system, 
over which one computes the Fermi speed $v_{F}(\boldsymbol{k}')=\left|\nabla_{\boldsymbol{k}'}E_{\nu}\right|$ over the $\nu$-th non-interacting band to perform the momentum integration. 
In this context, $\bar{\Gamma}(\boldsymbol{k},\boldsymbol{k}',\omega)$ contains the ground-state fluctuations and must, therefore, be computed at very low temperatures.

The quantity $F(\omega)$ is thus a measure of the strength of pairing potential and contains information regarding repulsive or attractive components of it. 
The vertex's imaginary part accounts for the momentum transferred after an interacting scattering event. 
Here, we aim to characterize the strength of the pairing interaction in a given channel so we use Eq.~\eqref{spectral_weights} to weigh over frequency, defining the pairing strength $\lambda$,
\begin{equation}
    \lambda=\int_{0}^{+\infty}d\omega\frac{F(\omega)}{\omega}.
\label{lbda_0}
\end{equation}

By making use of the Kramers-Kronig relation \cite{graserNeardegeneracySeveralPairing2009}, we obtain a frequency-independent form
\begin{equation}
    \lambda=-\sum_{\boldsymbol{k}'}\oint_{C_{j}}\frac{d\boldsymbol{k}'}{(2\pi)^{d}v_{F}(\boldsymbol{k}')}\text{Re}\bar{\Gamma}(\boldsymbol{k},\boldsymbol{k}',0).
\label{eq:lbda}
\end{equation}
The SC critical temperature is roughly given by $T_c=\omega_0 e^{-1/\lambda}$, where $\omega_0$ is a characteristic frequency cut-off of the spectral weights $F(\omega)$.

The parameter $\lambda$ defined in Eq.\eqref{eq:lbda} can be interpreted as an action that is minimized in some parameter region. Under the present definition, there is no upper boundary to the functional $\lambda$. Next, we naturally extend this idea to define the coupling strength functional that takes as argument the normalized gap symmetry $g(\boldsymbol{k})$ such that $\Delta(\boldsymbol{k})=\Delta_0 g(\boldsymbol{k})$, which is integrated over the $\boldsymbol{k}$ range,
\begin{equation}
    \begin{split}
        \lambda[g(\boldsymbol{k})]=-&\sum_{\boldsymbol{k}}\oint_{C_i } \frac{d\boldsymbol{k}}{(2\pi)^{d}v_{F}(\boldsymbol{k})}\sum_{\boldsymbol{k}'}\oint_{C_{j}}\frac{d\boldsymbol{k}'}{(2\pi)^{d}v_{F}(\boldsymbol{k}')} \\
        \times &g(\boldsymbol{k})\text{Re}\bar{\Gamma}(\boldsymbol{k},\boldsymbol{k}',0)g(\boldsymbol{k}') \\
        \times &\left[\sum_{\boldsymbol{k}'}\oint_{C_{j}}\frac{d\boldsymbol{k}'}{(2\pi)^{d}v_{F}(\boldsymbol{k}')}g(\boldsymbol{k}')^{2}\right]^{-1}.
    \end{split}
\label{lbda_func}
\end{equation}
Finally, we impose the stationary condition for this functional, i.e. $\delta\lambda/\delta g(\boldsymbol{k})=0$ to take into account mass renormalization \cite{qiPairingStrengthsTwo2008a}. This procedure results in the eigenvalues and eigenvectors Eq.~\eqref{gap} \cite{graserNeardegeneracySeveralPairing2009, scalapinoFermisurfaceInstabilitiesSuperconducting1987a}.
There, an index $\alpha$ labels the several solutions for the equation. However, the higher $\lambda_\alpha$ will generate the higher $T_c$, which is the physically realized critical temperature.
For example, under the constant density of states assumption, Eq.~\eqref{gap} retakes a BCS form
\begin{equation}
    -\sum_{\boldsymbol{k}'}\rho_{F}\bar{\Gamma}(\boldsymbol{k},\boldsymbol{k}')g_{\alpha}(\boldsymbol{k}')\approx\lambda_{\alpha} g_{\alpha}(\boldsymbol{k}).
\label{BCS}
\end{equation}

% ------------------------------

% ------------------------------
\section{Spin-triplet pairing vertex}
\label{appendix:triplet}

%-----------F I G U R E  ------
\begin{figure}[t]
\begin{center}
\includegraphics[width=1.0\columnwidth]{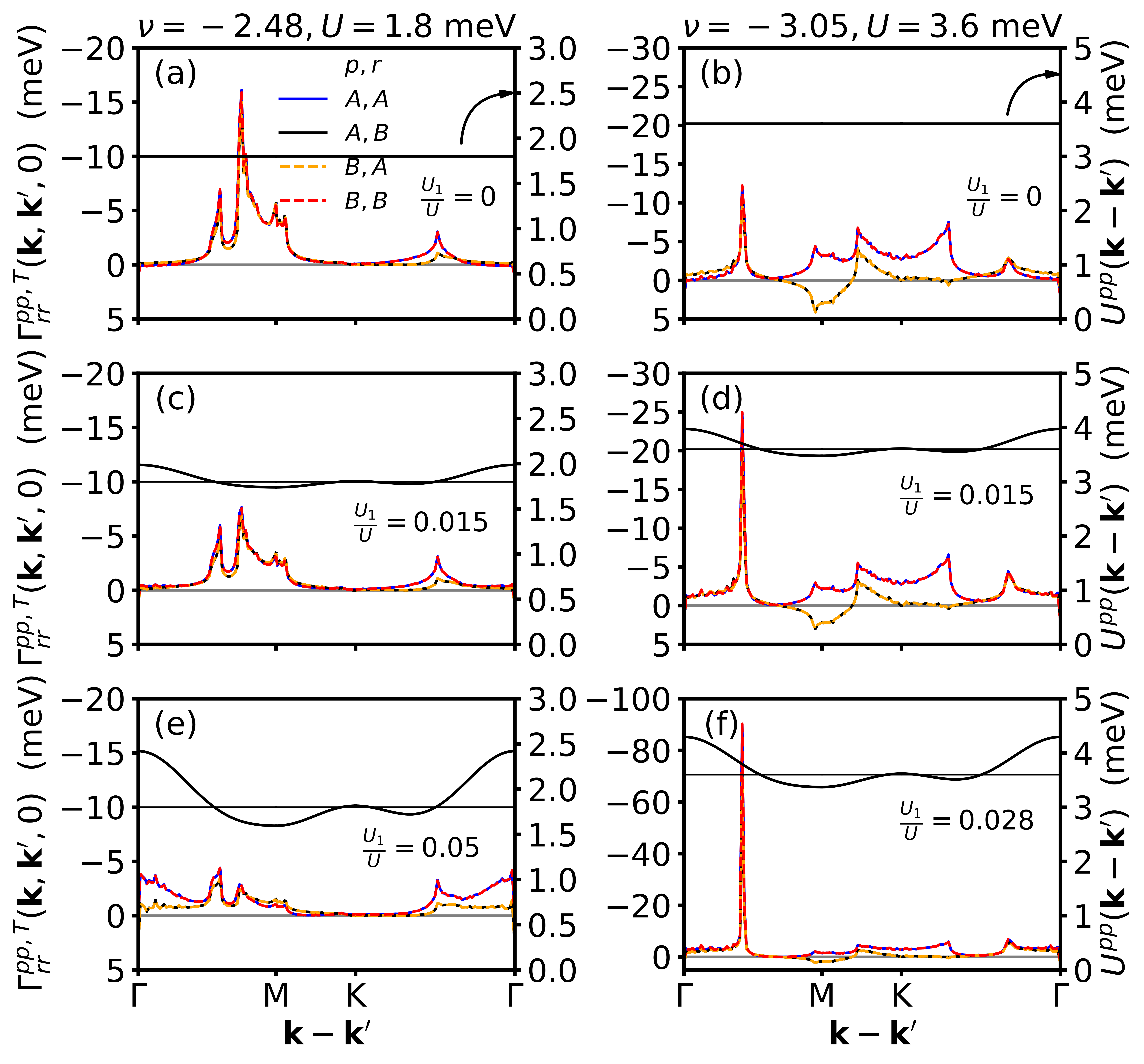}
\caption{
	Spin-triplet pairing vertex [Eq.~\eqref{triplet}] following the same notation as Fig.~\ref{fig:RPA_chi}. 
}
\label{fig:triplet}
\end{center}
\end{figure}
%-----------E N D  F I G U R E  ------

Fig.~\ref{fig:triplet} shows the spin-triplet pairing vertex defined as \cite{wuHarmonicFingerprintUnconventional2020}
\begin{align}
    \begin{split}
         \Gamma_{r\xi , t \xi^\prime}^{p \xi, q \xi^\prime, T}(\boldsymbol{k},\boldsymbol{k}',\omega)  & =\Bigg{[}-\frac{1}{2}\hat{U}_{s}\hat{\chi}_{s}(\boldsymbol{q}=\boldsymbol{k}-\boldsymbol{k}',\omega)\hat{U}_{s}+\frac{1}{2}\hat{U}_{s}  \\
	        & -\frac{1}{2}\hat{U}_{c}\hat{\chi}_{c}(\boldsymbol{q}=\boldsymbol{k}-\boldsymbol{k}',\omega)\hat{U}_{c}+\frac{1}{2}\hat{U}_{c}  \Bigg{]}_{p \xi, r\xi }^{ t \xi^\prime, q \xi^\prime}  \; . \label{triplet}
    \end{split}
\end{align}

As pointed out in the main text, the triplet channel has several attractive components but they are lower in magnitude than all the singlet channel ones [see the scale from -2 to 1 meV in Fig.~\ref{fig:triplet}(a) and 0 to 12 meV in Fig.~\ref{fig:RPA_chi}(a)]. For this reason, we neglect the triplet channel in our calculations of the pairing strength $\lambda$.

% ---------------------------------
\section{Weak dependence on the ratios $U_i/U_1$}
\label{appendix:ratios}

%-----------F I G U R E  - Appendix  ------
\begin{figure}[t]
\begin{center}
\includegraphics[width=1.0\columnwidth]{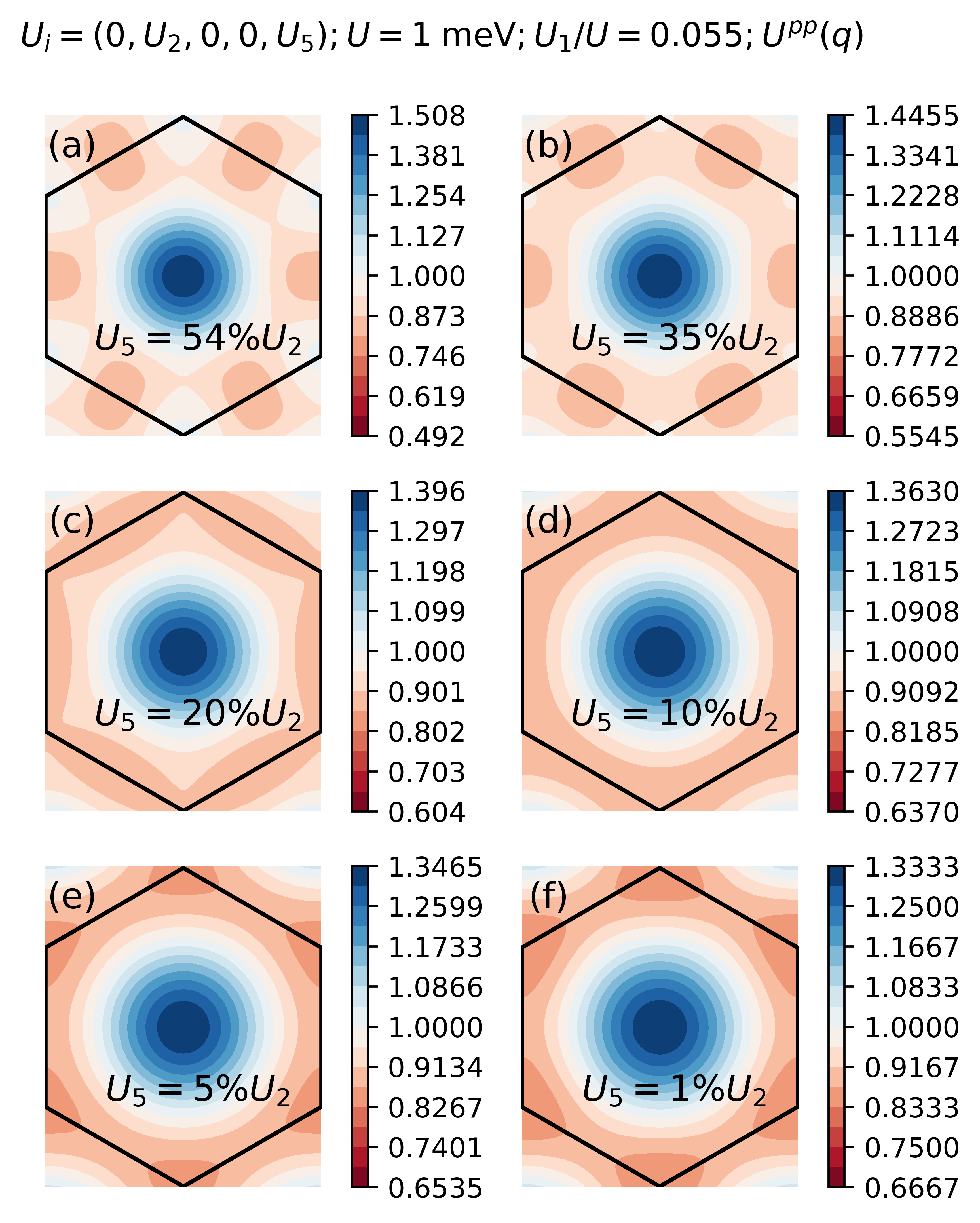}
\caption{  Same-sublattice Hubbard-like interactions $U^{pp}(\boldsymbol{q})$ (colors) for different fifth by second nearest-neighbor interaction strength ratios $U_5/U_2$. 
 }
\label{fig:appendix_Upp}
\end{center}
\end{figure}
%-----------E N D  F I G U R E   ------

In the main text, we showed simulations using the interaction ratios $U_i/U_1$ given in Ref.~\cite{koshinoMaximallyLocalizedWannier2018}, which used maximally localized Wannier functions to compute the interaction ratios.
These ratios can, however, depend on both the model and on the wannierization method used for their calculation. In this appendix, we show that our main results are robust, as they are not affected by small changes in these ratios.

As argued in Sec.~\ref{sec:pairingvertex}, the only relevant Hubbard-like interactions [Eq.~\eqref{eq:Uq}] for the spin fluctuations, which dominate pairing, are the same-sublattice elements $U^{pp}(\boldsymbol{q})$ with $p=A,B$. In addition, sublattice symmetry ensures that $U^{AA}(\boldsymbol{q})=U^{BB}(\boldsymbol{q})$.
Now, according to Fig.~\ref{fig:model}(a), the only same-sublattice neighbors are the second ($U_2$) and fifth ($U_5$), so the elements relevant for the superconducting instability are independent of other neighbors, such that we set them to zero in this appendix. Thus, a key ratio in our calculation is $U_5/U_2$, which is reported by Ref.~\cite{koshinoMaximallyLocalizedWannier2018}  as $0.5361$ (or $54\%$). 

As we argue in the main text, a key component of our results is the fact that $U^{pp}(\boldsymbol{q})$ displays minima around the $M$ points and that the position of the minima is independent of $U$ (see Fig.~\ref{fig:nestings}). In Fig.~\ref{fig:appendix_Upp}, we show that such minima 
are 
(i) robust upon a change in the ratio $U_5/U_2$, down to $35\%$ 
[panel (b)], and (ii) present for any finite long-range intrasublattice interaction strength.

Interestingly, for $U_5/U_2<35\%$, the minima in $U^{pp}(\boldsymbol{q})$ will develop around the $K$ points in scattering space, raising the question of which interaction minima cause the strongest pairing in MATBG, either $K$ or $M$ points.
Ref. \cite{calderonInteractions8orbitalModel2020} used an 8-orbital model fitted to a corrugated continuum model \cite{Carr:Phys.Rev.Res.:013001:2019} and found that interactions 
between neighbors separated by $L_M/r>50\%$ follow an $L_M/r$ curve. 
The second nearest neighbors (corresponding to the $U_2$ interaction strength) are separated by $L_M/r=1$, while the fifth nearest neighbors (corresponding to $U_5$) are separated by $L_M/r\approx2$.
These results give an estimate of the ratio 
$U_5/U_2\approx50\%$.
This is consistent with the result computed by Ref. \cite{koshinoMaximallyLocalizedWannier2018}, for which $U_5/U_2\approx54\%$ [Fig.~\ref{fig:appendix_Upp}(a)], establishing the minima at the $M$ points as the main model-independent feature originating the re-entrant magnetic state in MATBG as a function of the long-range interaction strength.

%\bibliography{TBG,TBG_v2,TBG_v3,Kagome}

\begin{thebibliography}{61}%
\makeatletter
\providecommand \@ifxundefined [1]{%
 \@ifx{#1\undefined}
}%
\providecommand \@ifnum [1]{%
 \ifnum #1\expandafter \@firstoftwo
 \else \expandafter \@secondoftwo
 \fi
}%
\providecommand \@ifx [1]{%
 \ifx #1\expandafter \@firstoftwo
 \else \expandafter \@secondoftwo
 \fi
}%
\providecommand \natexlab [1]{#1}%
\providecommand \enquote  [1]{``#1''}%
\providecommand \bibnamefont  [1]{#1}%
\providecommand \bibfnamefont [1]{#1}%
\providecommand \citenamefont [1]{#1}%
\providecommand \href@noop [0]{\@secondoftwo}%
\providecommand \href [0]{\begingroup \@sanitize@url \@href}%
\providecommand \@href[1]{\@@startlink{#1}\@@href}%
\providecommand \@@href[1]{\endgroup#1\@@endlink}%
\providecommand \@sanitize@url [0]{\catcode `\\12\catcode `\$12\catcode
  `\&12\catcode `\#12\catcode `\^12\catcode `\_12\catcode `\%12\relax}%
\providecommand \@@startlink[1]{}%
\providecommand \@@endlink[0]{}%
\providecommand \url  [0]{\begingroup\@sanitize@url \@url }%
\providecommand \@url [1]{\endgroup\@href {#1}{\urlprefix }}%
\providecommand \urlprefix  [0]{URL }%
\providecommand \Eprint [0]{\href }%
\providecommand \doibase [0]{https://doi.org/}%
\providecommand \selectlanguage [0]{\@gobble}%
\providecommand \bibinfo  [0]{\@secondoftwo}%
\providecommand \bibfield  [0]{\@secondoftwo}%
\providecommand \translation [1]{[#1]}%
\providecommand \BibitemOpen [0]{}%
\providecommand \bibitemStop [0]{}%
\providecommand \bibitemNoStop [0]{.\EOS\space}%
\providecommand \EOS [0]{\spacefactor3000\relax}%
\providecommand \BibitemShut  [1]{\csname bibitem#1\endcsname}%
\let\auto@bib@innerbib\@empty
%</preamble>
\bibitem [{\citenamefont {Cao}\ \emph {et~al.}(2018)\citenamefont {Cao},
  \citenamefont {Fatemi}, \citenamefont {Fang}, \citenamefont {Watanabe},
  \citenamefont {Taniguchi}, \citenamefont {Kaxiras},\ and\ \citenamefont
  {{Jarillo-Herrero}}}]{caoUnconventionalSuperconductivityMagicangle2018}%
  \BibitemOpen
  \bibfield  {author} {\bibinfo {author} {\bibfnamefont {Y.}~\bibnamefont
  {Cao}}, \bibinfo {author} {\bibfnamefont {V.}~\bibnamefont {Fatemi}},
  \bibinfo {author} {\bibfnamefont {S.}~\bibnamefont {Fang}}, \bibinfo {author}
  {\bibfnamefont {K.}~\bibnamefont {Watanabe}}, \bibinfo {author}
  {\bibfnamefont {T.}~\bibnamefont {Taniguchi}}, \bibinfo {author}
  {\bibfnamefont {E.}~\bibnamefont {Kaxiras}},\ and\ \bibinfo {author}
  {\bibfnamefont {P.}~\bibnamefont {{Jarillo-Herrero}}},\ }\bibfield  {title}
  {\bibinfo {title} {Unconventional superconductivity in magic-angle graphene
  superlattices},\ }\href {https://doi.org/10.1038/nature26160} {\bibfield
  {journal} {\bibinfo  {journal} {Nature}\ }\textbf {\bibinfo {volume} {556}},\
  \bibinfo {pages} {43} (\bibinfo {year} {2018})}\BibitemShut {NoStop}%
\bibitem [{\citenamefont {Polshyn}\ \emph {et~al.}(2019)\citenamefont
  {Polshyn}, \citenamefont {Yankowitz}, \citenamefont {Chen}, \citenamefont
  {Zhang}, \citenamefont {Watanabe}, \citenamefont {Taniguchi}, \citenamefont
  {Dean},\ and\ \citenamefont
  {Young}}]{polshynLargeLinearintemperatureResistivity2019}%
  \BibitemOpen
  \bibfield  {author} {\bibinfo {author} {\bibfnamefont {H.}~\bibnamefont
  {Polshyn}}, \bibinfo {author} {\bibfnamefont {M.}~\bibnamefont {Yankowitz}},
  \bibinfo {author} {\bibfnamefont {S.}~\bibnamefont {Chen}}, \bibinfo {author}
  {\bibfnamefont {Y.}~\bibnamefont {Zhang}}, \bibinfo {author} {\bibfnamefont
  {K.}~\bibnamefont {Watanabe}}, \bibinfo {author} {\bibfnamefont
  {T.}~\bibnamefont {Taniguchi}}, \bibinfo {author} {\bibfnamefont {C.~R.}\
  \bibnamefont {Dean}},\ and\ \bibinfo {author} {\bibfnamefont {A.~F.}\
  \bibnamefont {Young}},\ }\bibfield  {title} {\bibinfo {title} {Large
  linear-in-temperature resistivity in twisted bilayer graphene},\ }\href
  {https://doi.org/10.1038/s41567-019-0596-3} {\bibfield  {journal} {\bibinfo
  {journal} {Nat. Phys.}\ }\textbf {\bibinfo {volume} {15}},\ \bibinfo {pages}
  {1011} (\bibinfo {year} {2019})}\BibitemShut {NoStop}%
\bibitem [{\citenamefont {Cao}\ \emph {et~al.}(2020)\citenamefont {Cao},
  \citenamefont {Chowdhury}, \citenamefont {{Rodan-Legrain}}, \citenamefont
  {{Rubies-Bigorda}}, \citenamefont {Watanabe}, \citenamefont {Taniguchi},
  \citenamefont {Senthil},\ and\ \citenamefont
  {{Jarillo-Herrero}}}]{caoStrangeMetalMagicAngle2020}%
  \BibitemOpen
  \bibfield  {author} {\bibinfo {author} {\bibfnamefont {Y.}~\bibnamefont
  {Cao}}, \bibinfo {author} {\bibfnamefont {D.}~\bibnamefont {Chowdhury}},
  \bibinfo {author} {\bibfnamefont {D.}~\bibnamefont {{Rodan-Legrain}}},
  \bibinfo {author} {\bibfnamefont {O.}~\bibnamefont {{Rubies-Bigorda}}},
  \bibinfo {author} {\bibfnamefont {K.}~\bibnamefont {Watanabe}}, \bibinfo
  {author} {\bibfnamefont {T.}~\bibnamefont {Taniguchi}}, \bibinfo {author}
  {\bibfnamefont {T.}~\bibnamefont {Senthil}},\ and\ \bibinfo {author}
  {\bibfnamefont {P.}~\bibnamefont {{Jarillo-Herrero}}},\ }\bibfield  {title}
  {\bibinfo {title} {Strange {{Metal}} in {{Magic-Angle Graphene}} with near
  {{Planckian Dissipation}}},\ }\href
  {https://doi.org/10.1103/PhysRevLett.124.076801} {\bibfield  {journal}
  {\bibinfo  {journal} {Phys. Rev. Lett.}\ }\textbf {\bibinfo {volume} {124}},\
  \bibinfo {pages} {076801} (\bibinfo {year} {2020})}\BibitemShut {NoStop}%
\bibitem [{\citenamefont {Sharpe}\ \emph {et~al.}(2019)\citenamefont {Sharpe},
  \citenamefont {Fox}, \citenamefont {Barnard}, \citenamefont {Finney},
  \citenamefont {Watanabe}, \citenamefont {Taniguchi}, \citenamefont
  {Kastner},\ and\ \citenamefont
  {{Goldhaber-Gordon}}}]{sharpeEmergentFerromagnetismThreequarters2019}%
  \BibitemOpen
  \bibfield  {author} {\bibinfo {author} {\bibfnamefont {A.~L.}\ \bibnamefont
  {Sharpe}}, \bibinfo {author} {\bibfnamefont {E.~J.}\ \bibnamefont {Fox}},
  \bibinfo {author} {\bibfnamefont {A.~W.}\ \bibnamefont {Barnard}}, \bibinfo
  {author} {\bibfnamefont {J.}~\bibnamefont {Finney}}, \bibinfo {author}
  {\bibfnamefont {K.}~\bibnamefont {Watanabe}}, \bibinfo {author}
  {\bibfnamefont {T.}~\bibnamefont {Taniguchi}}, \bibinfo {author}
  {\bibfnamefont {M.~A.}\ \bibnamefont {Kastner}},\ and\ \bibinfo {author}
  {\bibfnamefont {D.}~\bibnamefont {{Goldhaber-Gordon}}},\ }\bibfield  {title}
  {\bibinfo {title} {Emergent ferromagnetism near three-quarters filling in
  twisted bilayer graphene},\ }\href {https://doi.org/10.1126/science.aaw3780}
  {\bibfield  {journal} {\bibinfo  {journal} {Science}\ }\textbf {\bibinfo
  {volume} {365}},\ \bibinfo {pages} {605} (\bibinfo {year}
  {2019})}\BibitemShut {NoStop}%
\bibitem [{\citenamefont {Serlin}\ \emph {et~al.}(2020)\citenamefont {Serlin},
  \citenamefont {Tschirhart}, \citenamefont {Polshyn}, \citenamefont {Zhang},
  \citenamefont {Zhu}, \citenamefont {Watanabe}, \citenamefont {Taniguchi},
  \citenamefont {Balents},\ and\ \citenamefont
  {Young}}]{serlinIntrinsicQuantizedAnomalous2020}%
  \BibitemOpen
  \bibfield  {author} {\bibinfo {author} {\bibfnamefont {M.}~\bibnamefont
  {Serlin}}, \bibinfo {author} {\bibfnamefont {C.~L.}\ \bibnamefont
  {Tschirhart}}, \bibinfo {author} {\bibfnamefont {H.}~\bibnamefont {Polshyn}},
  \bibinfo {author} {\bibfnamefont {Y.}~\bibnamefont {Zhang}}, \bibinfo
  {author} {\bibfnamefont {J.}~\bibnamefont {Zhu}}, \bibinfo {author}
  {\bibfnamefont {K.}~\bibnamefont {Watanabe}}, \bibinfo {author}
  {\bibfnamefont {T.}~\bibnamefont {Taniguchi}}, \bibinfo {author}
  {\bibfnamefont {L.}~\bibnamefont {Balents}},\ and\ \bibinfo {author}
  {\bibfnamefont {A.~F.}\ \bibnamefont {Young}},\ }\bibfield  {title} {\bibinfo
  {title} {Intrinsic quantized anomalous {{Hall}} effect in a moir\'e
  heterostructure},\ }\href {https://doi.org/10.1126/science.aay5533}
  {\bibfield  {journal} {\bibinfo  {journal} {Science}\ }\textbf {\bibinfo
  {volume} {367}},\ \bibinfo {pages} {900} (\bibinfo {year}
  {2020})}\BibitemShut {NoStop}%
\bibitem [{\citenamefont {Jiang}\ \emph {et~al.}(2019)\citenamefont {Jiang},
  \citenamefont {Lai}, \citenamefont {Watanabe}, \citenamefont {Taniguchi},
  \citenamefont {Haule}, \citenamefont {Mao},\ and\ \citenamefont
  {Andrei}}]{jiangChargeOrderBroken2019}%
  \BibitemOpen
  \bibfield  {author} {\bibinfo {author} {\bibfnamefont {Y.}~\bibnamefont
  {Jiang}}, \bibinfo {author} {\bibfnamefont {X.}~\bibnamefont {Lai}}, \bibinfo
  {author} {\bibfnamefont {K.}~\bibnamefont {Watanabe}}, \bibinfo {author}
  {\bibfnamefont {T.}~\bibnamefont {Taniguchi}}, \bibinfo {author}
  {\bibfnamefont {K.}~\bibnamefont {Haule}}, \bibinfo {author} {\bibfnamefont
  {J.}~\bibnamefont {Mao}},\ and\ \bibinfo {author} {\bibfnamefont {E.~Y.}\
  \bibnamefont {Andrei}},\ }\bibfield  {title} {\bibinfo {title} {Charge order
  and broken rotational symmetry in magic-angle twisted bilayer graphene},\
  }\href {https://doi.org/10.1038/s41586-019-1460-4} {\bibfield  {journal}
  {\bibinfo  {journal} {Nature}\ }\textbf {\bibinfo {volume} {573}},\ \bibinfo
  {pages} {91} (\bibinfo {year} {2019})}\BibitemShut {NoStop}%
\bibitem [{\citenamefont {Bhowmik}\ \emph {et~al.}(2022)\citenamefont
  {Bhowmik}, \citenamefont {Ghawri}, \citenamefont {Leconte}, \citenamefont
  {Appalakondaiah}, \citenamefont {Pandey}, \citenamefont {Mahapatra},
  \citenamefont {Lee}, \citenamefont {Watanabe}, \citenamefont {Taniguchi},
  \citenamefont {Jung}, \citenamefont {Ghosh},\ and\ \citenamefont
  {Chandni}}]{bhowmikBrokensymmetryStatesHalfinteger2022}%
  \BibitemOpen
  \bibfield  {author} {\bibinfo {author} {\bibfnamefont {S.}~\bibnamefont
  {Bhowmik}}, \bibinfo {author} {\bibfnamefont {B.}~\bibnamefont {Ghawri}},
  \bibinfo {author} {\bibfnamefont {N.}~\bibnamefont {Leconte}}, \bibinfo
  {author} {\bibfnamefont {S.}~\bibnamefont {Appalakondaiah}}, \bibinfo
  {author} {\bibfnamefont {M.}~\bibnamefont {Pandey}}, \bibinfo {author}
  {\bibfnamefont {P.~S.}\ \bibnamefont {Mahapatra}}, \bibinfo {author}
  {\bibfnamefont {D.}~\bibnamefont {Lee}}, \bibinfo {author} {\bibfnamefont
  {K.}~\bibnamefont {Watanabe}}, \bibinfo {author} {\bibfnamefont
  {T.}~\bibnamefont {Taniguchi}}, \bibinfo {author} {\bibfnamefont
  {J.}~\bibnamefont {Jung}}, \bibinfo {author} {\bibfnamefont {A.}~\bibnamefont
  {Ghosh}},\ and\ \bibinfo {author} {\bibfnamefont {U.}~\bibnamefont
  {Chandni}},\ }\bibfield  {title} {\bibinfo {title} {Broken-symmetry states at
  half-integer band fillings in twisted bilayer graphene},\ }\href
  {https://doi.org/10.1038/s41567-022-01557-4} {\bibfield  {journal} {\bibinfo
  {journal} {Nat. Phys.}\ }\textbf {\bibinfo {volume} {18}},\ \bibinfo {pages}
  {639} (\bibinfo {year} {2022})}\BibitemShut {NoStop}%
\bibitem [{\citenamefont {Choi}\ \emph {et~al.}(2019)\citenamefont {Choi},
  \citenamefont {Kemmer}, \citenamefont {Peng}, \citenamefont {Thomson},
  \citenamefont {Arora}, \citenamefont {Polski}, \citenamefont {Zhang},
  \citenamefont {Ren}, \citenamefont {Alicea}, \citenamefont {Refael},
  \citenamefont {{von Oppen}}, \citenamefont {Watanabe}, \citenamefont
  {Taniguchi},\ and\ \citenamefont
  {{Nadj-Perge}}}]{choiElectronicCorrelationsTwisted2019}%
  \BibitemOpen
  \bibfield  {author} {\bibinfo {author} {\bibfnamefont {Y.}~\bibnamefont
  {Choi}}, \bibinfo {author} {\bibfnamefont {J.}~\bibnamefont {Kemmer}},
  \bibinfo {author} {\bibfnamefont {Y.}~\bibnamefont {Peng}}, \bibinfo {author}
  {\bibfnamefont {A.}~\bibnamefont {Thomson}}, \bibinfo {author} {\bibfnamefont
  {H.}~\bibnamefont {Arora}}, \bibinfo {author} {\bibfnamefont
  {R.}~\bibnamefont {Polski}}, \bibinfo {author} {\bibfnamefont
  {Y.}~\bibnamefont {Zhang}}, \bibinfo {author} {\bibfnamefont
  {H.}~\bibnamefont {Ren}}, \bibinfo {author} {\bibfnamefont {J.}~\bibnamefont
  {Alicea}}, \bibinfo {author} {\bibfnamefont {G.}~\bibnamefont {Refael}},
  \bibinfo {author} {\bibfnamefont {F.}~\bibnamefont {{von Oppen}}}, \bibinfo
  {author} {\bibfnamefont {K.}~\bibnamefont {Watanabe}}, \bibinfo {author}
  {\bibfnamefont {T.}~\bibnamefont {Taniguchi}},\ and\ \bibinfo {author}
  {\bibfnamefont {S.}~\bibnamefont {{Nadj-Perge}}},\ }\bibfield  {title}
  {\bibinfo {title} {Electronic correlations in twisted bilayer graphene near
  the magic angle},\ }\href {https://doi.org/10.1038/s41567-019-0606-5}
  {\bibfield  {journal} {\bibinfo  {journal} {Nat. Phys.}\ }\textbf {\bibinfo
  {volume} {15}},\ \bibinfo {pages} {1174} (\bibinfo {year}
  {2019})}\BibitemShut {NoStop}%
\bibitem [{\citenamefont {Balents}\ \emph {et~al.}(2020)\citenamefont
  {Balents}, \citenamefont {Dean}, \citenamefont {Efetov},\ and\ \citenamefont
  {Young}}]{balentsSuperconductivityStrongCorrelations2020}%
  \BibitemOpen
  \bibfield  {author} {\bibinfo {author} {\bibfnamefont {L.}~\bibnamefont
  {Balents}}, \bibinfo {author} {\bibfnamefont {C.~R.}\ \bibnamefont {Dean}},
  \bibinfo {author} {\bibfnamefont {D.~K.}\ \bibnamefont {Efetov}},\ and\
  \bibinfo {author} {\bibfnamefont {A.~F.}\ \bibnamefont {Young}},\ }\bibfield
  {title} {\bibinfo {title} {Superconductivity and strong correlations in
  moir\'e flat bands},\ }\href {https://doi.org/10.1038/s41567-020-0906-9}
  {\bibfield  {journal} {\bibinfo  {journal} {Nat. Phys.}\ }\textbf {\bibinfo
  {volume} {16}},\ \bibinfo {pages} {725} (\bibinfo {year} {2020})}\BibitemShut
  {NoStop}%
\bibitem [{\citenamefont {Cao}\ \emph {et~al.}(2021)\citenamefont {Cao},
  \citenamefont {{Rodan-Legrain}}, \citenamefont {Park}, \citenamefont {Yuan},
  \citenamefont {Watanabe}, \citenamefont {Taniguchi}, \citenamefont
  {Fernandes}, \citenamefont {Fu},\ and\ \citenamefont
  {{Jarillo-Herrero}}}]{caoNematicityCompetingOrders2021}%
  \BibitemOpen
  \bibfield  {author} {\bibinfo {author} {\bibfnamefont {Y.}~\bibnamefont
  {Cao}}, \bibinfo {author} {\bibfnamefont {D.}~\bibnamefont
  {{Rodan-Legrain}}}, \bibinfo {author} {\bibfnamefont {J.~M.}\ \bibnamefont
  {Park}}, \bibinfo {author} {\bibfnamefont {N.~F.~Q.}\ \bibnamefont {Yuan}},
  \bibinfo {author} {\bibfnamefont {K.}~\bibnamefont {Watanabe}}, \bibinfo
  {author} {\bibfnamefont {T.}~\bibnamefont {Taniguchi}}, \bibinfo {author}
  {\bibfnamefont {R.~M.}\ \bibnamefont {Fernandes}}, \bibinfo {author}
  {\bibfnamefont {L.}~\bibnamefont {Fu}},\ and\ \bibinfo {author}
  {\bibfnamefont {P.}~\bibnamefont {{Jarillo-Herrero}}},\ }\bibfield  {title}
  {\bibinfo {title} {Nematicity and competing orders in superconducting
  magic-angle graphene},\ }\href {https://doi.org/10.1126/science.abc2836}
  {\bibfield  {journal} {\bibinfo  {journal} {Science}\ }\textbf {\bibinfo
  {volume} {372}},\ \bibinfo {pages} {264} (\bibinfo {year}
  {2021})}\BibitemShut {NoStop}%
\bibitem [{\citenamefont {Nuckolls}\ \emph {et~al.}(2020)\citenamefont
  {Nuckolls}, \citenamefont {Oh}, \citenamefont {Wong}, \citenamefont {Lian},
  \citenamefont {Watanabe}, \citenamefont {Taniguchi}, \citenamefont
  {Bernevig},\ and\ \citenamefont
  {Yazdani}}]{nuckollsStronglyCorrelatedChern2020}%
  \BibitemOpen
  \bibfield  {author} {\bibinfo {author} {\bibfnamefont {K.~P.}\ \bibnamefont
  {Nuckolls}}, \bibinfo {author} {\bibfnamefont {M.}~\bibnamefont {Oh}},
  \bibinfo {author} {\bibfnamefont {D.}~\bibnamefont {Wong}}, \bibinfo {author}
  {\bibfnamefont {B.}~\bibnamefont {Lian}}, \bibinfo {author} {\bibfnamefont
  {K.}~\bibnamefont {Watanabe}}, \bibinfo {author} {\bibfnamefont
  {T.}~\bibnamefont {Taniguchi}}, \bibinfo {author} {\bibfnamefont {B.~A.}\
  \bibnamefont {Bernevig}},\ and\ \bibinfo {author} {\bibfnamefont
  {A.}~\bibnamefont {Yazdani}},\ }\bibfield  {title} {\bibinfo {title}
  {Strongly correlated {{Chern}} insulators in magic-angle twisted bilayer
  graphene},\ }\href {https://doi.org/10.1038/s41586-020-3028-8} {\bibfield
  {journal} {\bibinfo  {journal} {Nature}\ }\textbf {\bibinfo {volume} {588}},\
  \bibinfo {pages} {610} (\bibinfo {year} {2020})}\BibitemShut {NoStop}%
\bibitem [{\citenamefont {Roy}\ and\ \citenamefont {Juri\ifmmode \check{c}\else
  \v{c}\fi{}i\ifmmode~\acute{c}\else
  \'{c}\fi{}}(2019)}]{Roy:Phys.Rev.B:121407:2019}%
  \BibitemOpen
  \bibfield  {author} {\bibinfo {author} {\bibfnamefont {B.}~\bibnamefont
  {Roy}}\ and\ \bibinfo {author} {\bibfnamefont {V.}~\bibnamefont {Juri\ifmmode
  \check{c}\else \v{c}\fi{}i\ifmmode~\acute{c}\else \'{c}\fi{}}},\ }\bibfield
  {title} {\bibinfo {title} {Unconventional superconductivity in nearly flat
  bands in twisted bilayer graphene},\ }\href
  {https://doi.org/10.1103/PhysRevB.99.121407} {\bibfield  {journal} {\bibinfo
  {journal} {Phys. Rev. B}\ }\textbf {\bibinfo {volume} {99}},\ \bibinfo
  {pages} {121407(R)} (\bibinfo {year} {2019})}\BibitemShut {NoStop}%
\bibitem [{\citenamefont {Scalapino}(2012)}]{Scalapino2012}%
  \BibitemOpen
  \bibfield  {author} {\bibinfo {author} {\bibfnamefont {D.~J.}\ \bibnamefont
  {Scalapino}},\ }\bibfield  {title} {\bibinfo {title} {{A common thread: The
  pairing interaction for unconventional superconductors}},\ }\href
  {https://doi.org/10.1103/RevModPhys.84.1383} {\bibfield  {journal} {\bibinfo
  {journal} {Rev. Mod. Phys.}\ }\textbf {\bibinfo {volume} {84}},\ \bibinfo
  {pages} {1383} (\bibinfo {year} {2012})}\BibitemShut {NoStop}%
\bibitem [{\citenamefont {Graser}\ \emph {et~al.}(2009)\citenamefont {Graser},
  \citenamefont {Maier}, \citenamefont {Hirschfeld},\ and\ \citenamefont
  {ino}}]{graserNeardegeneracySeveralPairing2009}%
  \BibitemOpen
  \bibfield  {author} {\bibinfo {author} {\bibfnamefont {S.}~\bibnamefont
  {Graser}}, \bibinfo {author} {\bibfnamefont {T.~A.}\ \bibnamefont {Maier}},
  \bibinfo {author} {\bibfnamefont {P.~J.}\ \bibnamefont {Hirschfeld}},\ and\
  \bibinfo {author} {\bibfnamefont {D.~J.}\ \bibnamefont {ino}},\ }\bibfield
  {title} {\bibinfo {title} {Near-degeneracy of several pairing channels in
  multiorbital models for the {{Fe}} pnictides},\ }\href
  {https://doi.org/10.1088/1367-2630/11/2/025016} {\bibfield  {journal}
  {\bibinfo  {journal} {New J. Phys.}\ }\textbf {\bibinfo {volume} {11}},\
  \bibinfo {pages} {025016} (\bibinfo {year} {2009})}\BibitemShut {NoStop}%
\bibitem [{\citenamefont {Kerelsky}\ \emph {et~al.}(2019)\citenamefont
  {Kerelsky}, \citenamefont {McGilly}, \citenamefont {Kennes}, \citenamefont
  {Xian}, \citenamefont {Yankowitz}, \citenamefont {Chen}, \citenamefont
  {Watanabe}, \citenamefont {Taniguchi}, \citenamefont {Hone}, \citenamefont
  {Dean}, \citenamefont {Rubio},\ and\ \citenamefont
  {Pasupathy}}]{kerelskyMaximizedElectronInteractions2019}%
  \BibitemOpen
  \bibfield  {author} {\bibinfo {author} {\bibfnamefont {A.}~\bibnamefont
  {Kerelsky}}, \bibinfo {author} {\bibfnamefont {L.~J.}\ \bibnamefont
  {McGilly}}, \bibinfo {author} {\bibfnamefont {D.~M.}\ \bibnamefont {Kennes}},
  \bibinfo {author} {\bibfnamefont {L.}~\bibnamefont {Xian}}, \bibinfo {author}
  {\bibfnamefont {M.}~\bibnamefont {Yankowitz}}, \bibinfo {author}
  {\bibfnamefont {S.}~\bibnamefont {Chen}}, \bibinfo {author} {\bibfnamefont
  {K.}~\bibnamefont {Watanabe}}, \bibinfo {author} {\bibfnamefont
  {T.}~\bibnamefont {Taniguchi}}, \bibinfo {author} {\bibfnamefont
  {J.}~\bibnamefont {Hone}}, \bibinfo {author} {\bibfnamefont {C.}~\bibnamefont
  {Dean}}, \bibinfo {author} {\bibfnamefont {A.}~\bibnamefont {Rubio}},\ and\
  \bibinfo {author} {\bibfnamefont {A.~N.}\ \bibnamefont {Pasupathy}},\
  }\bibfield  {title} {\bibinfo {title} {Maximized electron interactions at the
  magic angle in twisted bilayer graphene},\ }\href
  {https://doi.org/10.1038/s41586-019-1431-9} {\bibfield  {journal} {\bibinfo
  {journal} {Nature}\ }\textbf {\bibinfo {volume} {572}},\ \bibinfo {pages}
  {95} (\bibinfo {year} {2019})}\BibitemShut {NoStop}%
\bibitem [{\citenamefont {Liu}\ \emph {et~al.}(2021)\citenamefont {Liu},
  \citenamefont {Wang}, \citenamefont {Watanabe}, \citenamefont {Taniguchi},
  \citenamefont {Vafek},\ and\ \citenamefont
  {Li}}]{liuTuningElectronCorrelation2021}%
  \BibitemOpen
  \bibfield  {author} {\bibinfo {author} {\bibfnamefont {X.}~\bibnamefont
  {Liu}}, \bibinfo {author} {\bibfnamefont {Z.}~\bibnamefont {Wang}}, \bibinfo
  {author} {\bibfnamefont {K.}~\bibnamefont {Watanabe}}, \bibinfo {author}
  {\bibfnamefont {T.}~\bibnamefont {Taniguchi}}, \bibinfo {author}
  {\bibfnamefont {O.}~\bibnamefont {Vafek}},\ and\ \bibinfo {author}
  {\bibfnamefont {J.~I.~A.}\ \bibnamefont {Li}},\ }\bibfield  {title} {\bibinfo
  {title} {Tuning electron correlation in magic-angle twisted bilayer graphene
  using {{Coulomb}} screening},\ }\href
  {https://doi.org/10.1126/science.abb8754} {\bibfield  {journal} {\bibinfo
  {journal} {Science}\ }\textbf {\bibinfo {volume} {371}},\ \bibinfo {pages}
  {1261} (\bibinfo {year} {2021})}\BibitemShut {NoStop}%
\bibitem [{\citenamefont {Lian}\ \emph {et~al.}(2019)\citenamefont {Lian},
  \citenamefont {Wang},\ and\ \citenamefont {Bernevig}}]{Lian2019}%
  \BibitemOpen
  \bibfield  {author} {\bibinfo {author} {\bibfnamefont {B.}~\bibnamefont
  {Lian}}, \bibinfo {author} {\bibfnamefont {Z.}~\bibnamefont {Wang}},\ and\
  \bibinfo {author} {\bibfnamefont {B.~A.}\ \bibnamefont {Bernevig}},\
  }\bibfield  {title} {\bibinfo {title} {Twisted bilayer graphene: A
  phonon-driven superconductor},\ }\href
  {https://doi.org/10.1103/PhysRevLett.122.257002} {\bibfield  {journal}
  {\bibinfo  {journal} {Phys. Rev. Lett.}\ }\textbf {\bibinfo {volume} {122}},\
  \bibinfo {pages} {257002} (\bibinfo {year} {2019})}\BibitemShut {NoStop}%
\bibitem [{Note1()}]{Note1}%
  \BibitemOpen
  \bibinfo {note} {In this paper, we refer to Hubbard-like interactions beyond
  nearest-neighbors as ``long-range'', although we are aware that some authors,
  e.g. Ref.~\cite {defenuLongrangeInteractingQuantum2023}, prefer to use the
  term ``non-local'' for such terms.}\BibitemShut {Stop}%
\bibitem [{\citenamefont {Koshino}\ \emph {et~al.}(2018)\citenamefont
  {Koshino}, \citenamefont {Yuan}, \citenamefont {Koretsune}, \citenamefont
  {Ochi}, \citenamefont {Kuroki},\ and\ \citenamefont
  {Fu}}]{koshinoMaximallyLocalizedWannier2018}%
  \BibitemOpen
  \bibfield  {author} {\bibinfo {author} {\bibfnamefont {M.}~\bibnamefont
  {Koshino}}, \bibinfo {author} {\bibfnamefont {N.~F.~Q.}\ \bibnamefont
  {Yuan}}, \bibinfo {author} {\bibfnamefont {T.}~\bibnamefont {Koretsune}},
  \bibinfo {author} {\bibfnamefont {M.}~\bibnamefont {Ochi}}, \bibinfo {author}
  {\bibfnamefont {K.}~\bibnamefont {Kuroki}},\ and\ \bibinfo {author}
  {\bibfnamefont {L.}~\bibnamefont {Fu}},\ }\bibfield  {title} {\bibinfo
  {title} {Maximally {{Localized Wannier Orbitals}} and the {{Extended Hubbard
  Model}} for {{Twisted Bilayer Graphene}}},\ }\href
  {https://doi.org/10.1103/PhysRevX.8.031087} {\bibfield  {journal} {\bibinfo
  {journal} {Phys. Rev. X}\ }\textbf {\bibinfo {volume} {8}},\ \bibinfo {pages}
  {031087} (\bibinfo {year} {2018})}\BibitemShut {NoStop}%
\bibitem [{\citenamefont {Bistritzer}\ and\ \citenamefont
  {MacDonald}(2011)}]{bistritzerMoireBandsTwisted2011}%
  \BibitemOpen
  \bibfield  {author} {\bibinfo {author} {\bibfnamefont {R.}~\bibnamefont
  {Bistritzer}}\ and\ \bibinfo {author} {\bibfnamefont {A.~H.}\ \bibnamefont
  {MacDonald}},\ }\bibfield  {title} {\bibinfo {title} {Moir\'e bands in
  twisted double-layer graphene},\ }\href
  {https://doi.org/10.1073/pnas.1108174108} {\bibfield  {journal} {\bibinfo
  {journal} {PNAS}\ }\textbf {\bibinfo {volume} {108}},\ \bibinfo {pages}
  {12233} (\bibinfo {year} {2011})}\BibitemShut {NoStop}%
\bibitem [{\citenamefont {Lucignano}\ \emph {et~al.}(2019)\citenamefont
  {Lucignano}, \citenamefont {Alf{\`e}}, \citenamefont {Cataudella},
  \citenamefont {Ninno},\ and\ \citenamefont
  {Cantele}}]{lucignanoCrucialRoleAtomic2019a}%
  \BibitemOpen
  \bibfield  {author} {\bibinfo {author} {\bibfnamefont {P.}~\bibnamefont
  {Lucignano}}, \bibinfo {author} {\bibfnamefont {D.}~\bibnamefont {Alf{\`e}}},
  \bibinfo {author} {\bibfnamefont {V.}~\bibnamefont {Cataudella}}, \bibinfo
  {author} {\bibfnamefont {D.}~\bibnamefont {Ninno}},\ and\ \bibinfo {author}
  {\bibfnamefont {G.}~\bibnamefont {Cantele}},\ }\bibfield  {title} {\bibinfo
  {title} {Crucial role of atomic corrugation on the flat bands and energy gaps
  of twisted bilayer graphene at the magic angle {$\theta\sim 1.08^\circ$}},\
  }\href {https://doi.org/10.1103/PhysRevB.99.195419} {\bibfield  {journal}
  {\bibinfo  {journal} {Phys. Rev. B}\ }\textbf {\bibinfo {volume} {99}},\
  \bibinfo {pages} {195419} (\bibinfo {year} {2019})}\BibitemShut {NoStop}%
\bibitem [{\citenamefont {Esirgen}\ \emph {et~al.}(1999)\citenamefont
  {Esirgen}, \citenamefont {Sch{\"u}ttler},\ and\ \citenamefont
  {Bickers}}]{esirgenMathitWavePairing1999a}%
  \BibitemOpen
  \bibfield  {author} {\bibinfo {author} {\bibfnamefont {G.}~\bibnamefont
  {Esirgen}}, \bibinfo {author} {\bibfnamefont {H.-B.}\ \bibnamefont
  {Sch{\"u}ttler}},\ and\ \bibinfo {author} {\bibfnamefont {N.~E.}\
  \bibnamefont {Bickers}},\ }\bibfield  {title} {\bibinfo {title} {{$d$}-{{Wave
  Pairing}} in the {{Presence}} of {{Long-Range Coulomb Interactions}}},\
  }\href {https://doi.org/10.1103/PhysRevLett.82.1217} {\bibfield  {journal}
  {\bibinfo  {journal} {Phys. Rev. Lett.}\ }\textbf {\bibinfo {volume} {82}},\
  \bibinfo {pages} {1217} (\bibinfo {year} {1999})}\BibitemShut {NoStop}%
\bibitem [{\citenamefont {Altmeyer}\ \emph {et~al.}(2016)\citenamefont
  {Altmeyer}, \citenamefont {Guterding}, \citenamefont {Hirschfeld},
  \citenamefont {Maier}, \citenamefont {Valent{\'i}},\ and\ \citenamefont
  {Scalapino}}]{altmeyerRoleVertexCorrections2016}%
  \BibitemOpen
  \bibfield  {author} {\bibinfo {author} {\bibfnamefont {M.}~\bibnamefont
  {Altmeyer}}, \bibinfo {author} {\bibfnamefont {D.}~\bibnamefont {Guterding}},
  \bibinfo {author} {\bibfnamefont {P.~J.}\ \bibnamefont {Hirschfeld}},
  \bibinfo {author} {\bibfnamefont {T.~A.}\ \bibnamefont {Maier}}, \bibinfo
  {author} {\bibfnamefont {R.}~\bibnamefont {Valent{\'i}}},\ and\ \bibinfo
  {author} {\bibfnamefont {D.~J.}\ \bibnamefont {Scalapino}},\ }\bibfield
  {title} {\bibinfo {title} {Role of vertex corrections in the matrix
  formulation of the random phase approximation for the multiorbital
  {{Hubbard}} model},\ }\href {https://doi.org/10.1103/PhysRevB.94.214515}
  {\bibfield  {journal} {\bibinfo  {journal} {Phys. Rev. B}\ }\textbf {\bibinfo
  {volume} {94}},\ \bibinfo {pages} {214515} (\bibinfo {year}
  {2016})}\BibitemShut {NoStop}%
\bibitem [{\citenamefont {Wu}\ and\ \citenamefont
  {Das~Sarma}(2019)}]{wuIdentificationSuperconductingPairing2019}%
  \BibitemOpen
  \bibfield  {author} {\bibinfo {author} {\bibfnamefont {F.}~\bibnamefont
  {Wu}}\ and\ \bibinfo {author} {\bibfnamefont {S.}~\bibnamefont {Das~Sarma}},\
  }\bibfield  {title} {\bibinfo {title} {Identification of superconducting
  pairing symmetry in twisted bilayer graphene using in-plane magnetic field
  and strain},\ }\href {https://doi.org/10.1103/PhysRevB.99.220507} {\bibfield
  {journal} {\bibinfo  {journal} {Phys. Rev. B}\ }\textbf {\bibinfo {volume}
  {99}},\ \bibinfo {pages} {220507(R)} (\bibinfo {year} {2019})}\BibitemShut
  {NoStop}%
\bibitem [{\citenamefont {T{\"o}rm{\"a}}\ \emph {et~al.}(2022)\citenamefont
  {T{\"o}rm{\"a}}, \citenamefont {Peotta},\ and\ \citenamefont
  {Bernevig}}]{tormaSuperconductivitySuperfluidityQuantum2022a}%
  \BibitemOpen
  \bibfield  {author} {\bibinfo {author} {\bibfnamefont {P.}~\bibnamefont
  {T{\"o}rm{\"a}}}, \bibinfo {author} {\bibfnamefont {S.}~\bibnamefont
  {Peotta}},\ and\ \bibinfo {author} {\bibfnamefont {B.~A.}\ \bibnamefont
  {Bernevig}},\ }\bibfield  {title} {\bibinfo {title} {Superconductivity,
  superfluidity and quantum geometry in twisted multilayer systems},\ }\href
  {https://doi.org/10.1038/s42254-022-00466-y} {\bibfield  {journal} {\bibinfo
  {journal} {Nat. Rev. Phys.}\ }\textbf {\bibinfo {volume} {4}},\ \bibinfo
  {pages} {528} (\bibinfo {year} {2022})}\BibitemShut {NoStop}%
\bibitem [{\citenamefont {Lopes~dos Santos}\ \emph {et~al.}(2007)\citenamefont
  {Lopes~dos Santos}, \citenamefont {Peres},\ and\ \citenamefont
  {Castro~Neto}}]{LopesdosSantos:Phys.Rev.Lett.:256802:2007}%
  \BibitemOpen
  \bibfield  {author} {\bibinfo {author} {\bibfnamefont {J.~M.~B.}\
  \bibnamefont {Lopes~dos Santos}}, \bibinfo {author} {\bibfnamefont
  {N.~M.~R.}\ \bibnamefont {Peres}},\ and\ \bibinfo {author} {\bibfnamefont
  {A.~H.}\ \bibnamefont {Castro~Neto}},\ }\bibfield  {title} {\bibinfo {title}
  {Graphene bilayer with a twist: Electronic structure},\ }\href
  {https://doi.org/10.1103/PhysRevLett.99.256802} {\bibfield  {journal}
  {\bibinfo  {journal} {Phys. Rev. Lett.}\ }\textbf {\bibinfo {volume} {99}},\
  \bibinfo {pages} {256802} (\bibinfo {year} {2007})}\BibitemShut {NoStop}%
\bibitem [{\citenamefont {L{\"o}thman}\ \emph {et~al.}(2022)\citenamefont
  {L{\"o}thman}, \citenamefont {Schmidt}, \citenamefont {Parhizgar},\ and\
  \citenamefont
  {{Black-Schaffer}}}]{lothmanNematicSuperconductivityMagicangle2022}%
  \BibitemOpen
  \bibfield  {author} {\bibinfo {author} {\bibfnamefont {T.}~\bibnamefont
  {L{\"o}thman}}, \bibinfo {author} {\bibfnamefont {J.}~\bibnamefont
  {Schmidt}}, \bibinfo {author} {\bibfnamefont {F.}~\bibnamefont {Parhizgar}},\
  and\ \bibinfo {author} {\bibfnamefont {A.~M.}\ \bibnamefont
  {{Black-Schaffer}}},\ }\bibfield  {title} {\bibinfo {title} {Nematic
  superconductivity in magic-angle twisted bilayer graphene from atomistic
  modeling},\ }\href {https://doi.org/10.1038/s42005-022-00860-z} {\bibfield
  {journal} {\bibinfo  {journal} {Comm. Phys.}\ }\textbf {\bibinfo {volume}
  {5}},\ \bibinfo {pages} {1} (\bibinfo {year} {2022})}\BibitemShut {NoStop}%
\bibitem [{\citenamefont {Kang}\ and\ \citenamefont
  {Vafek}(2018)}]{kangSymmetryMaximallyLocalized2018}%
  \BibitemOpen
  \bibfield  {author} {\bibinfo {author} {\bibfnamefont {J.}~\bibnamefont
  {Kang}}\ and\ \bibinfo {author} {\bibfnamefont {O.}~\bibnamefont {Vafek}},\
  }\bibfield  {title} {\bibinfo {title} {Symmetry, {{Maximally Localized
  Wannier States}}, and a {{Low-Energy Model}} for {{Twisted Bilayer Graphene
  Narrow Bands}}},\ }\href {https://doi.org/10.1103/PhysRevX.8.031088}
  {\bibfield  {journal} {\bibinfo  {journal} {Phys. Rev. X}\ }\textbf {\bibinfo
  {volume} {8}},\ \bibinfo {pages} {031088} (\bibinfo {year}
  {2018})}\BibitemShut {NoStop}%
\bibitem [{\citenamefont {Carr}\ \emph {et~al.}(2019)\citenamefont {Carr},
  \citenamefont {Fang}, \citenamefont {Zhu},\ and\ \citenamefont
  {Kaxiras}}]{Carr:Phys.Rev.Res.:013001:2019}%
  \BibitemOpen
  \bibfield  {author} {\bibinfo {author} {\bibfnamefont {S.}~\bibnamefont
  {Carr}}, \bibinfo {author} {\bibfnamefont {S.}~\bibnamefont {Fang}}, \bibinfo
  {author} {\bibfnamefont {Z.}~\bibnamefont {Zhu}},\ and\ \bibinfo {author}
  {\bibfnamefont {E.}~\bibnamefont {Kaxiras}},\ }\bibfield  {title} {\bibinfo
  {title} {Exact continuum model for low-energy electronic states of twisted
  bilayer graphene},\ }\href {https://doi.org/10.1103/PhysRevResearch.1.013001}
  {\bibfield  {journal} {\bibinfo  {journal} {Phys. Rev. Res.}\ }\textbf
  {\bibinfo {volume} {1}},\ \bibinfo {pages} {013001} (\bibinfo {year}
  {2019})}\BibitemShut {NoStop}%
\bibitem [{\citenamefont {Zou}\ \emph {et~al.}(2018)\citenamefont {Zou},
  \citenamefont {Po}, \citenamefont {Vishwanath},\ and\ \citenamefont
  {Senthil}}]{Zou:Phys.Rev.B:085435:2018}%
  \BibitemOpen
  \bibfield  {author} {\bibinfo {author} {\bibfnamefont {L.}~\bibnamefont
  {Zou}}, \bibinfo {author} {\bibfnamefont {H.~C.}\ \bibnamefont {Po}},
  \bibinfo {author} {\bibfnamefont {A.}~\bibnamefont {Vishwanath}},\ and\
  \bibinfo {author} {\bibfnamefont {T.}~\bibnamefont {Senthil}},\ }\bibfield
  {title} {\bibinfo {title} {Band structure of twisted bilayer graphene:
  Emergent symmetries, commensurate approximants, and wannier obstructions},\
  }\href {https://doi.org/10.1103/PhysRevB.98.085435} {\bibfield  {journal}
  {\bibinfo  {journal} {Phys. Rev. B}\ }\textbf {\bibinfo {volume} {98}},\
  \bibinfo {pages} {085435} (\bibinfo {year} {2018})}\BibitemShut {NoStop}%
\bibitem [{\citenamefont {Po}\ \emph {et~al.}(2018)\citenamefont {Po},
  \citenamefont {Zou}, \citenamefont {Vishwanath},\ and\ \citenamefont
  {Senthil}}]{Po:Phys.Rev.X:031089:2018}%
  \BibitemOpen
  \bibfield  {author} {\bibinfo {author} {\bibfnamefont {H.~C.}\ \bibnamefont
  {Po}}, \bibinfo {author} {\bibfnamefont {L.}~\bibnamefont {Zou}}, \bibinfo
  {author} {\bibfnamefont {A.}~\bibnamefont {Vishwanath}},\ and\ \bibinfo
  {author} {\bibfnamefont {T.}~\bibnamefont {Senthil}},\ }\bibfield  {title}
  {\bibinfo {title} {Origin of mott insulating behavior and superconductivity
  in twisted bilayer graphene},\ }\href
  {https://doi.org/10.1103/PhysRevX.8.031089} {\bibfield  {journal} {\bibinfo
  {journal} {Phys. Rev. X}\ }\textbf {\bibinfo {volume} {8}},\ \bibinfo {pages}
  {031089} (\bibinfo {year} {2018})}\BibitemShut {NoStop}%
\bibitem [{\citenamefont {Po}\ \emph {et~al.}(2019)\citenamefont {Po},
  \citenamefont {Zou}, \citenamefont {Senthil},\ and\ \citenamefont
  {Vishwanath}}]{poFaithfulTightbindingModels2019a}%
  \BibitemOpen
  \bibfield  {author} {\bibinfo {author} {\bibfnamefont {H.~C.}\ \bibnamefont
  {Po}}, \bibinfo {author} {\bibfnamefont {L.}~\bibnamefont {Zou}}, \bibinfo
  {author} {\bibfnamefont {T.}~\bibnamefont {Senthil}},\ and\ \bibinfo {author}
  {\bibfnamefont {A.}~\bibnamefont {Vishwanath}},\ }\bibfield  {title}
  {\bibinfo {title} {Faithful tight-binding models and fragile topology of
  magic-angle bilayer graphene},\ }\href
  {https://doi.org/10.1103/PhysRevB.99.195455} {\bibfield  {journal} {\bibinfo
  {journal} {Phys. Rev. B}\ }\textbf {\bibinfo {volume} {99}},\ \bibinfo
  {pages} {195455} (\bibinfo {year} {2019})}\BibitemShut {NoStop}%
\bibitem [{\citenamefont {Song}\ \emph {et~al.}(2019)\citenamefont {Song},
  \citenamefont {Wang}, \citenamefont {Shi}, \citenamefont {Li}, \citenamefont
  {Fang},\ and\ \citenamefont {Bernevig}}]{songAllMagicAngles2019}%
  \BibitemOpen
  \bibfield  {author} {\bibinfo {author} {\bibfnamefont {Z.}~\bibnamefont
  {Song}}, \bibinfo {author} {\bibfnamefont {Z.}~\bibnamefont {Wang}}, \bibinfo
  {author} {\bibfnamefont {W.}~\bibnamefont {Shi}}, \bibinfo {author}
  {\bibfnamefont {G.}~\bibnamefont {Li}}, \bibinfo {author} {\bibfnamefont
  {C.}~\bibnamefont {Fang}},\ and\ \bibinfo {author} {\bibfnamefont {B.~A.}\
  \bibnamefont {Bernevig}},\ }\bibfield  {title} {\bibinfo {title} {All magic
  angles in twisted bilayer graphene are topological},\ }\href
  {https://doi.org/10.1103/PhysRevLett.123.036401} {\bibfield  {journal}
  {\bibinfo  {journal} {Phys. Rev. Lett.}\ }\textbf {\bibinfo {volume} {123}},\
  \bibinfo {pages} {036401} (\bibinfo {year} {2019})}\BibitemShut {NoStop}%
\bibitem [{\citenamefont {Calder{\'o}n}\ and\ \citenamefont
  {Bascones}(2020)}]{calderonInteractions8orbitalModel2020}%
  \BibitemOpen
  \bibfield  {author} {\bibinfo {author} {\bibfnamefont {M.~J.}\ \bibnamefont
  {Calder{\'o}n}}\ and\ \bibinfo {author} {\bibfnamefont {E.}~\bibnamefont
  {Bascones}},\ }\bibfield  {title} {\bibinfo {title} {Interactions in the
  8-orbital model for twisted bilayer graphene},\ }\href
  {https://doi.org/10.1103/PhysRevB.102.155149} {\bibfield  {journal} {\bibinfo
   {journal} {Phys. Rev. B}\ }\textbf {\bibinfo {volume} {102}},\ \bibinfo
  {pages} {155149} (\bibinfo {year} {2020})}\BibitemShut {NoStop}%
\bibitem [{\citenamefont {Wu}\ \emph {et~al.}(2020)\citenamefont {Wu},
  \citenamefont {Hanke}, \citenamefont {Fink}, \citenamefont {Klett},\ and\
  \citenamefont {Thomale}}]{wuHarmonicFingerprintUnconventional2020}%
  \BibitemOpen
  \bibfield  {author} {\bibinfo {author} {\bibfnamefont {X.}~\bibnamefont
  {Wu}}, \bibinfo {author} {\bibfnamefont {W.}~\bibnamefont {Hanke}}, \bibinfo
  {author} {\bibfnamefont {M.}~\bibnamefont {Fink}}, \bibinfo {author}
  {\bibfnamefont {M.}~\bibnamefont {Klett}},\ and\ \bibinfo {author}
  {\bibfnamefont {R.}~\bibnamefont {Thomale}},\ }\bibfield  {title} {\bibinfo
  {title} {Harmonic fingerprint of unconventional superconductivity in twisted
  bilayer graphene},\ }\href {https://doi.org/10.1103/PhysRevB.101.134517}
  {\bibfield  {journal} {\bibinfo  {journal} {Phys. Rev. B}\ }\textbf {\bibinfo
  {volume} {101}},\ \bibinfo {pages} {134517} (\bibinfo {year}
  {2020})}\BibitemShut {NoStop}%
\bibitem [{\citenamefont {Pizarro}\ \emph {et~al.}(2019)\citenamefont
  {Pizarro}, \citenamefont {R{\"o}sner}, \citenamefont {Thomale}, \citenamefont
  {Valent{\'i}},\ and\ \citenamefont
  {Wehling}}]{pizarroInternalScreeningDielectric2019}%
  \BibitemOpen
  \bibfield  {author} {\bibinfo {author} {\bibfnamefont {J.~M.}\ \bibnamefont
  {Pizarro}}, \bibinfo {author} {\bibfnamefont {M.}~\bibnamefont {R{\"o}sner}},
  \bibinfo {author} {\bibfnamefont {R.}~\bibnamefont {Thomale}}, \bibinfo
  {author} {\bibfnamefont {R.}~\bibnamefont {Valent{\'i}}},\ and\ \bibinfo
  {author} {\bibfnamefont {T.~O.}\ \bibnamefont {Wehling}},\ }\bibfield
  {title} {\bibinfo {title} {Internal screening and dielectric engineering in
  magic-angle twisted bilayer graphene},\ }\href
  {https://doi.org/10.1103/PhysRevB.100.161102} {\bibfield  {journal} {\bibinfo
   {journal} {Phys. Rev. B}\ }\textbf {\bibinfo {volume} {100}},\ \bibinfo
  {pages} {161102(R)} (\bibinfo {year} {2019})}\BibitemShut {NoStop}%
\bibitem [{\citenamefont {Goodwin}\ \emph {et~al.}(2019)\citenamefont
  {Goodwin}, \citenamefont {Corsetti}, \citenamefont {Mostofi},\ and\
  \citenamefont
  {Lischner}}]{goodwinAttractiveElectronelectronInteractions2019}%
  \BibitemOpen
  \bibfield  {author} {\bibinfo {author} {\bibfnamefont {Z.~A.~H.}\
  \bibnamefont {Goodwin}}, \bibinfo {author} {\bibfnamefont {F.}~\bibnamefont
  {Corsetti}}, \bibinfo {author} {\bibfnamefont {A.~A.}\ \bibnamefont
  {Mostofi}},\ and\ \bibinfo {author} {\bibfnamefont {J.}~\bibnamefont
  {Lischner}},\ }\bibfield  {title} {\bibinfo {title} {Attractive
  electron-electron interactions from internal screening in magic-angle twisted
  bilayer graphene},\ }\href {https://doi.org/10.1103/PhysRevB.100.235424}
  {\bibfield  {journal} {\bibinfo  {journal} {Phys. Rev. B}\ }\textbf {\bibinfo
  {volume} {100}},\ \bibinfo {pages} {235424} (\bibinfo {year}
  {2019})}\BibitemShut {NoStop}%
\bibitem [{\citenamefont {Vanhala}\ and\ \citenamefont
  {Pollet}(2020)}]{vanhalaConstrainedRandomPhase2020}%
  \BibitemOpen
  \bibfield  {author} {\bibinfo {author} {\bibfnamefont {T.~I.}\ \bibnamefont
  {Vanhala}}\ and\ \bibinfo {author} {\bibfnamefont {L.}~\bibnamefont
  {Pollet}},\ }\bibfield  {title} {\bibinfo {title} {Constrained random phase
  approximation of the effective {{Coulomb}} interaction in lattice models of
  twisted bilayer graphene},\ }\href
  {https://doi.org/10.1103/PhysRevB.102.035154} {\bibfield  {journal} {\bibinfo
   {journal} {Phys. Rev. B}\ }\textbf {\bibinfo {volume} {102}},\ \bibinfo
  {pages} {035154} (\bibinfo {year} {2020})}\BibitemShut {NoStop}%
\bibitem [{\citenamefont {Leconte}\ \emph {et~al.}(2022)\citenamefont
  {Leconte}, \citenamefont {Javvaji}, \citenamefont {An}, \citenamefont
  {Samudrala},\ and\ \citenamefont
  {Jung}}]{leconteRelaxationEffectsTwisted2022}%
  \BibitemOpen
  \bibfield  {author} {\bibinfo {author} {\bibfnamefont {N.}~\bibnamefont
  {Leconte}}, \bibinfo {author} {\bibfnamefont {S.}~\bibnamefont {Javvaji}},
  \bibinfo {author} {\bibfnamefont {J.}~\bibnamefont {An}}, \bibinfo {author}
  {\bibfnamefont {A.}~\bibnamefont {Samudrala}},\ and\ \bibinfo {author}
  {\bibfnamefont {J.}~\bibnamefont {Jung}},\ }\bibfield  {title} {\bibinfo
  {title} {Relaxation effects in twisted bilayer graphene: {{A}} multiscale
  approach},\ }\href {https://doi.org/10.1103/PhysRevB.106.115410} {\bibfield
  {journal} {\bibinfo  {journal} {Phys. Rev. B}\ }\textbf {\bibinfo {volume}
  {106}},\ \bibinfo {pages} {115410} (\bibinfo {year} {2022})}\BibitemShut
  {NoStop}%
\bibitem [{\citenamefont {Liu}\ \emph {et~al.}(2018)\citenamefont {Liu},
  \citenamefont {Zhang}, \citenamefont {Chen},\ and\ \citenamefont
  {Yang}}]{liuChiralSpinDensity2018}%
  \BibitemOpen
  \bibfield  {author} {\bibinfo {author} {\bibfnamefont {C.-C.}\ \bibnamefont
  {Liu}}, \bibinfo {author} {\bibfnamefont {L.-D.}\ \bibnamefont {Zhang}},
  \bibinfo {author} {\bibfnamefont {W.-Q.}\ \bibnamefont {Chen}},\ and\
  \bibinfo {author} {\bibfnamefont {F.}~\bibnamefont {Yang}},\ }\bibfield
  {title} {\bibinfo {title} {Chiral {{Spin Density Wave}} and {$d+id$}
  {{Superconductivity}} in the {{Magic-Angle-Twisted Bilayer Graphene}}},\
  }\href {https://doi.org/10.1103/PhysRevLett.121.217001} {\bibfield  {journal}
  {\bibinfo  {journal} {Phys. Rev. Lett.}\ }\textbf {\bibinfo {volume} {121}},\
  \bibinfo {pages} {217001} (\bibinfo {year} {2018})}\BibitemShut {NoStop}%
\bibitem [{\citenamefont {Huang}\ \emph
  {et~al.}(2022{\natexlab{a}})\citenamefont {Huang}, \citenamefont {Tu},
  \citenamefont {Shen}, \citenamefont {Zheng}, \citenamefont {Wang},
  \citenamefont {Wang}, \citenamefont {Khaliji}, \citenamefont {Park},
  \citenamefont {Liu}, \citenamefont {Yang}, \citenamefont {Zhang},
  \citenamefont {Shao}, \citenamefont {Li}, \citenamefont {Low}, \citenamefont
  {Shi},\ and\ \citenamefont {Wang}}]{huangObservationChiralSlow2022}%
  \BibitemOpen
  \bibfield  {author} {\bibinfo {author} {\bibfnamefont {T.}~\bibnamefont
  {Huang}}, \bibinfo {author} {\bibfnamefont {X.}~\bibnamefont {Tu}}, \bibinfo
  {author} {\bibfnamefont {C.}~\bibnamefont {Shen}}, \bibinfo {author}
  {\bibfnamefont {B.}~\bibnamefont {Zheng}}, \bibinfo {author} {\bibfnamefont
  {J.}~\bibnamefont {Wang}}, \bibinfo {author} {\bibfnamefont {H.}~\bibnamefont
  {Wang}}, \bibinfo {author} {\bibfnamefont {K.}~\bibnamefont {Khaliji}},
  \bibinfo {author} {\bibfnamefont {S.~H.}\ \bibnamefont {Park}}, \bibinfo
  {author} {\bibfnamefont {Z.}~\bibnamefont {Liu}}, \bibinfo {author}
  {\bibfnamefont {T.}~\bibnamefont {Yang}}, \bibinfo {author} {\bibfnamefont
  {Z.}~\bibnamefont {Zhang}}, \bibinfo {author} {\bibfnamefont
  {L.}~\bibnamefont {Shao}}, \bibinfo {author} {\bibfnamefont {X.}~\bibnamefont
  {Li}}, \bibinfo {author} {\bibfnamefont {T.}~\bibnamefont {Low}}, \bibinfo
  {author} {\bibfnamefont {Y.}~\bibnamefont {Shi}},\ and\ \bibinfo {author}
  {\bibfnamefont {X.}~\bibnamefont {Wang}},\ }\bibfield  {title} {\bibinfo
  {title} {Observation of chiral and slow plasmons in twisted bilayer
  graphene},\ }\href {https://doi.org/10.1038/s41586-022-04520-8} {\bibfield
  {journal} {\bibinfo  {journal} {Nature}\ }\textbf {\bibinfo {volume} {605}},\
  \bibinfo {pages} {63} (\bibinfo {year} {2022}{\natexlab{a}})}\BibitemShut
  {NoStop}%
\bibitem [{\citenamefont {Huang}\ \emph
  {et~al.}(2022{\natexlab{b}})\citenamefont {Huang}, \citenamefont {Wei},
  \citenamefont {Qin},\ and\ \citenamefont
  {MacDonald}}]{huangPseudospinParamagnonsSuperconducting2022}%
  \BibitemOpen
  \bibfield  {author} {\bibinfo {author} {\bibfnamefont {C.}~\bibnamefont
  {Huang}}, \bibinfo {author} {\bibfnamefont {N.}~\bibnamefont {Wei}}, \bibinfo
  {author} {\bibfnamefont {W.}~\bibnamefont {Qin}},\ and\ \bibinfo {author}
  {\bibfnamefont {A.~H.}\ \bibnamefont {MacDonald}},\ }\bibfield  {title}
  {\bibinfo {title} {Pseudospin {{Paramagnons}} and the {{Superconducting
  Dome}} in {{Magic Angle Twisted Bilayer Graphene}}},\ }\href
  {https://doi.org/10.1103/PhysRevLett.129.187001} {\bibfield  {journal}
  {\bibinfo  {journal} {Phys. Rev. Lett.}\ }\textbf {\bibinfo {volume} {129}},\
  \bibinfo {pages} {187001} (\bibinfo {year} {2022}{\natexlab{b}})}\BibitemShut
  {NoStop}%
\bibitem [{\citenamefont {Luo}\ \emph {et~al.}(2010)\citenamefont {Luo},
  \citenamefont {Martins}, \citenamefont {Yao}, \citenamefont {Daghofer},
  \citenamefont {Yu}, \citenamefont {Moreo},\ and\ \citenamefont
  {Dagotto}}]{luoNeutronARPESConstraints2010}%
  \BibitemOpen
  \bibfield  {author} {\bibinfo {author} {\bibfnamefont {Q.}~\bibnamefont
  {Luo}}, \bibinfo {author} {\bibfnamefont {G.}~\bibnamefont {Martins}},
  \bibinfo {author} {\bibfnamefont {D.-X.}\ \bibnamefont {Yao}}, \bibinfo
  {author} {\bibfnamefont {M.}~\bibnamefont {Daghofer}}, \bibinfo {author}
  {\bibfnamefont {R.}~\bibnamefont {Yu}}, \bibinfo {author} {\bibfnamefont
  {A.}~\bibnamefont {Moreo}},\ and\ \bibinfo {author} {\bibfnamefont
  {E.}~\bibnamefont {Dagotto}},\ }\bibfield  {title} {\bibinfo {title} {Neutron
  and {{ARPES}} constraints on the couplings of the multiorbital {{Hubbard}}
  model for the iron pnictides},\ }\href
  {https://doi.org/10.1103/PhysRevB.82.104508} {\bibfield  {journal} {\bibinfo
  {journal} {Phys. Rev. B}\ }\textbf {\bibinfo {volume} {82}},\ \bibinfo
  {pages} {104508} (\bibinfo {year} {2010})}\BibitemShut {NoStop}%
\bibitem [{\citenamefont {Nicholson}\ \emph
  {et~al.}(2011{\natexlab{a}})\citenamefont {Nicholson}, \citenamefont {Luo},
  \citenamefont {Ge}, \citenamefont {Riera}, \citenamefont {Daghofer},
  \citenamefont {Martins}, \citenamefont {Moreo},\ and\ \citenamefont
  {Dagotto}}]{nicholsonRoleDegeneracyHybridization2011}%
  \BibitemOpen
  \bibfield  {author} {\bibinfo {author} {\bibfnamefont {A.}~\bibnamefont
  {Nicholson}}, \bibinfo {author} {\bibfnamefont {Q.}~\bibnamefont {Luo}},
  \bibinfo {author} {\bibfnamefont {W.}~\bibnamefont {Ge}}, \bibinfo {author}
  {\bibfnamefont {J.}~\bibnamefont {Riera}}, \bibinfo {author} {\bibfnamefont
  {M.}~\bibnamefont {Daghofer}}, \bibinfo {author} {\bibfnamefont {G.~B.}\
  \bibnamefont {Martins}}, \bibinfo {author} {\bibfnamefont {A.}~\bibnamefont
  {Moreo}},\ and\ \bibinfo {author} {\bibfnamefont {E.}~\bibnamefont
  {Dagotto}},\ }\bibfield  {title} {\bibinfo {title} {Role of degeneracy,
  hybridization, and nesting in the properties of multiorbital systems},\
  }\href {https://doi.org/10.1103/PhysRevB.84.094519} {\bibfield  {journal}
  {\bibinfo  {journal} {Phys. Rev. B}\ }\textbf {\bibinfo {volume} {84}},\
  \bibinfo {pages} {094519} (\bibinfo {year} {2011}{\natexlab{a}})}\BibitemShut
  {NoStop}%
\bibitem [{\citenamefont {Nicholson}\ \emph
  {et~al.}(2011{\natexlab{b}})\citenamefont {Nicholson}, \citenamefont {Ge},
  \citenamefont {Zhang}, \citenamefont {Riera}, \citenamefont {Daghofer},
  \citenamefont {Ole{\'s}}, \citenamefont {Martins}, \citenamefont {Moreo},\
  and\ \citenamefont {Dagotto}}]{nicholsonCompetingPairingSymmetries2011}%
  \BibitemOpen
  \bibfield  {author} {\bibinfo {author} {\bibfnamefont {A.}~\bibnamefont
  {Nicholson}}, \bibinfo {author} {\bibfnamefont {W.}~\bibnamefont {Ge}},
  \bibinfo {author} {\bibfnamefont {X.}~\bibnamefont {Zhang}}, \bibinfo
  {author} {\bibfnamefont {J.}~\bibnamefont {Riera}}, \bibinfo {author}
  {\bibfnamefont {M.}~\bibnamefont {Daghofer}}, \bibinfo {author}
  {\bibfnamefont {A.~M.}\ \bibnamefont {Ole{\'s}}}, \bibinfo {author}
  {\bibfnamefont {G.~B.}\ \bibnamefont {Martins}}, \bibinfo {author}
  {\bibfnamefont {A.}~\bibnamefont {Moreo}},\ and\ \bibinfo {author}
  {\bibfnamefont {E.}~\bibnamefont {Dagotto}},\ }\bibfield  {title} {\bibinfo
  {title} {Competing {{Pairing Symmetries}} in a {{Generalized Two-Orbital
  Model}} for the {{Pnictide Superconductors}}},\ }\href
  {https://doi.org/10.1103/PhysRevLett.106.217002} {\bibfield  {journal}
  {\bibinfo  {journal} {Phys. Rev. Lett.}\ }\textbf {\bibinfo {volume} {106}},\
  \bibinfo {pages} {217002} (\bibinfo {year} {2011}{\natexlab{b}})}\BibitemShut
  {NoStop}%
\bibitem [{\citenamefont {Martins}\ \emph {et~al.}(2013)\citenamefont
  {Martins}, \citenamefont {Moreo},\ and\ \citenamefont
  {Dagotto}}]{martinsRPAAnalysisTwoorbital2013a}%
  \BibitemOpen
  \bibfield  {author} {\bibinfo {author} {\bibfnamefont {G.~B.}\ \bibnamefont
  {Martins}}, \bibinfo {author} {\bibfnamefont {A.}~\bibnamefont {Moreo}},\
  and\ \bibinfo {author} {\bibfnamefont {E.}~\bibnamefont {Dagotto}},\
  }\bibfield  {title} {\bibinfo {title} {{{RPA}} analysis of a two-orbital
  model for the {{BiS}}\$\{\}\_\{2\}\$-based superconductors},\ }\href
  {https://doi.org/10.1103/PhysRevB.87.081102} {\bibfield  {journal} {\bibinfo
  {journal} {Phys. Rev. B}\ }\textbf {\bibinfo {volume} {87}},\ \bibinfo
  {pages} {081102(R)} (\bibinfo {year} {2013})}\BibitemShut {NoStop}%
\bibitem [{Note2()}]{Note2}%
  \BibitemOpen
  \bibinfo {note} {A more detailed derivation of the interaction matrices is
  presented in Ref.~\cite {brazChargeSpinFluctuations2024}.}\BibitemShut
  {Stop}%
\bibitem [{Note3()}]{Note3}%
  \BibitemOpen
  \bibinfo {note} {We note that our main results hold over a range of values
  for the $U_i/U$ ratios. For instance, while the ratios of Ref.~\cite
  {koshinoMaximallyLocalizedWannier2018} result in $U_5/U_2=0.5362$, we find
  that the results of Figs.~\ref {fig:lambda_filling} and \ref {fig:RPA_chi}
  will hold even for $U_5/U_2 \sim 0.35$. See Appendix \ref {appendix:ratios}
  for a full discussion.}\BibitemShut {Stop}%
\bibitem [{\citenamefont {Fischer}\ \emph {et~al.}(2021)\citenamefont
  {Fischer}, \citenamefont {Klebl}, \citenamefont {Honerkamp},\ and\
  \citenamefont {Kennes}}]{fischerSpinfluctuationinducedPairingTwisted2021}%
  \BibitemOpen
  \bibfield  {author} {\bibinfo {author} {\bibfnamefont {A.}~\bibnamefont
  {Fischer}}, \bibinfo {author} {\bibfnamefont {L.}~\bibnamefont {Klebl}},
  \bibinfo {author} {\bibfnamefont {C.}~\bibnamefont {Honerkamp}},\ and\
  \bibinfo {author} {\bibfnamefont {D.~M.}\ \bibnamefont {Kennes}},\ }\bibfield
   {title} {\bibinfo {title} {Spin-fluctuation-induced pairing in twisted
  bilayer graphene},\ }\href {https://doi.org/10.1103/PhysRevB.103.L041103}
  {\bibfield  {journal} {\bibinfo  {journal} {Phys. Rev. B}\ }\textbf {\bibinfo
  {volume} {103}},\ \bibinfo {pages} {L041103} (\bibinfo {year}
  {2021})}\BibitemShut {NoStop}%
\bibitem [{\citenamefont {Zhang}\ \emph {et~al.}(2020)\citenamefont {Zhang},
  \citenamefont {Zhang}, \citenamefont {Lu}, \citenamefont {Chen},\ and\
  \citenamefont {Yang}}]{zhangDensityWaveTopological2020}%
  \BibitemOpen
  \bibfield  {author} {\bibinfo {author} {\bibfnamefont {M.}~\bibnamefont
  {Zhang}}, \bibinfo {author} {\bibfnamefont {Y.}~\bibnamefont {Zhang}},
  \bibinfo {author} {\bibfnamefont {C.}~\bibnamefont {Lu}}, \bibinfo {author}
  {\bibfnamefont {W.-Q.}\ \bibnamefont {Chen}},\ and\ \bibinfo {author}
  {\bibfnamefont {F.}~\bibnamefont {Yang}},\ }\bibfield  {title} {\bibinfo
  {title} {Density wave and topological superconductivity in the
  magic-angle-twisted bilayer-graphene*},\ }\href
  {https://doi.org/10.1088/1674-1056/abc7b5} {\bibfield  {journal} {\bibinfo
  {journal} {Chin. Phys. B}\ }\textbf {\bibinfo {volume} {29}},\ \bibinfo
  {pages} {127102} (\bibinfo {year} {2020})}\BibitemShut {NoStop}%
\bibitem [{\citenamefont {Takimoto}\ \emph {et~al.}(2004)\citenamefont
  {Takimoto}, \citenamefont {Hotta},\ and\ \citenamefont
  {Ueda}}]{takimotoStrongcouplingTheorySuperconductivity2004}%
  \BibitemOpen
  \bibfield  {author} {\bibinfo {author} {\bibfnamefont {T.}~\bibnamefont
  {Takimoto}}, \bibinfo {author} {\bibfnamefont {T.}~\bibnamefont {Hotta}},\
  and\ \bibinfo {author} {\bibfnamefont {K.}~\bibnamefont {Ueda}},\ }\bibfield
  {title} {\bibinfo {title} {Strong-coupling theory of superconductivity in a
  degenerate {{Hubbard}} model},\ }\href
  {https://doi.org/10.1103/PhysRevB.69.104504} {\bibfield  {journal} {\bibinfo
  {journal} {Phys. Rev. B}\ }\textbf {\bibinfo {volume} {69}},\ \bibinfo
  {pages} {104504} (\bibinfo {year} {2004})}\BibitemShut {NoStop}%
\bibitem [{\citenamefont {Berk}\ and\ \citenamefont
  {Schrieffer}(1966)}]{berkEffectFerromagneticSpin1966a}%
  \BibitemOpen
  \bibfield  {author} {\bibinfo {author} {\bibfnamefont {N.~F.}\ \bibnamefont
  {Berk}}\ and\ \bibinfo {author} {\bibfnamefont {J.~R.}\ \bibnamefont
  {Schrieffer}},\ }\bibfield  {title} {\bibinfo {title} {Effect of
  {{Ferromagnetic Spin Correlations}} on {{Superconductivity}}},\ }\href
  {https://doi.org/10.1103/PhysRevLett.17.433} {\bibfield  {journal} {\bibinfo
  {journal} {Phys. Rev. Lett.}\ }\textbf {\bibinfo {volume} {17}},\ \bibinfo
  {pages} {433} (\bibinfo {year} {1966})}\BibitemShut {NoStop}%
\bibitem [{\citenamefont {Bickers}\ and\ \citenamefont
  {Scalapino}(1989)}]{bickersConservingApproximationsStrongly1989b}%
  \BibitemOpen
  \bibfield  {author} {\bibinfo {author} {\bibfnamefont {N.~E.}\ \bibnamefont
  {Bickers}}\ and\ \bibinfo {author} {\bibfnamefont {D.~J.}\ \bibnamefont
  {Scalapino}},\ }\bibfield  {title} {\bibinfo {title} {Conserving
  approximations for strongly fluctuating electron systems. {{I}}.
  {{Formalism}} and calculational approach},\ }\href
  {https://doi.org/10.1016/0003-4916(89)90359-X} {\bibfield  {journal}
  {\bibinfo  {journal} {Ann. Phys.}\ }\textbf {\bibinfo {volume} {193}},\
  \bibinfo {pages} {206} (\bibinfo {year} {1989})}\BibitemShut {NoStop}%
\bibitem [{\citenamefont {Bickers}\ and\ \citenamefont
  {White}(1991)}]{bickersConservingApproximationsStrongly1991}%
  \BibitemOpen
  \bibfield  {author} {\bibinfo {author} {\bibfnamefont {N.~E.}\ \bibnamefont
  {Bickers}}\ and\ \bibinfo {author} {\bibfnamefont {S.~R.}\ \bibnamefont
  {White}},\ }\bibfield  {title} {\bibinfo {title} {Conserving approximations
  for strongly fluctuating electron systems. {{II}}. {{Numerical}} results and
  parquet extension},\ }\href {https://doi.org/10.1103/PhysRevB.43.8044}
  {\bibfield  {journal} {\bibinfo  {journal} {Phys. Rev. B}\ }\textbf {\bibinfo
  {volume} {43}},\ \bibinfo {pages} {8044} (\bibinfo {year}
  {1991})}\BibitemShut {NoStop}%
\bibitem [{\citenamefont {Sakakibara}\ \emph {et~al.}(2012)\citenamefont
  {Sakakibara}, \citenamefont {Usui}, \citenamefont {Kuroki}, \citenamefont
  {Arita},\ and\ \citenamefont
  {Aoki}}]{sakakibaraOriginMaterialDependence2012}%
  \BibitemOpen
  \bibfield  {author} {\bibinfo {author} {\bibfnamefont {H.}~\bibnamefont
  {Sakakibara}}, \bibinfo {author} {\bibfnamefont {H.}~\bibnamefont {Usui}},
  \bibinfo {author} {\bibfnamefont {K.}~\bibnamefont {Kuroki}}, \bibinfo
  {author} {\bibfnamefont {R.}~\bibnamefont {Arita}},\ and\ \bibinfo {author}
  {\bibfnamefont {H.}~\bibnamefont {Aoki}},\ }\bibfield  {title} {\bibinfo
  {title} {Origin of the material dependence of $t_c$ in the single-layered
  cuprates},\ }\href {https://doi.org/10.1103/PhysRevB.85.064501} {\bibfield
  {journal} {\bibinfo  {journal} {Phys. Rev. B}\ }\textbf {\bibinfo {volume}
  {85}},\ \bibinfo {pages} {064501} (\bibinfo {year} {2012})}\BibitemShut
  {NoStop}%
\bibitem [{\citenamefont {Engstr\"om}\ \emph {et~al.}(2023)\citenamefont
  {Engstr\"om}, \citenamefont {Liu}, \citenamefont {Witczak-Krempa},\ and\
  \citenamefont {Pereg-Barnea}}]{Engstroem:Phys.Rev.B:014508:2023}%
  \BibitemOpen
  \bibfield  {author} {\bibinfo {author} {\bibfnamefont {L.}~\bibnamefont
  {Engstr\"om}}, \bibinfo {author} {\bibfnamefont {C.-C.}\ \bibnamefont {Liu}},
  \bibinfo {author} {\bibfnamefont {W.}~\bibnamefont {Witczak-Krempa}},\ and\
  \bibinfo {author} {\bibfnamefont {T.}~\bibnamefont {Pereg-Barnea}},\
  }\bibfield  {title} {\bibinfo {title} {Strain-induced superconductivity in
  ${\mathrm{sr}}_{2}{\mathrm{iro}}_{4}$},\ }\href
  {https://doi.org/10.1103/PhysRevB.108.014508} {\bibfield  {journal} {\bibinfo
   {journal} {Phys. Rev. B}\ }\textbf {\bibinfo {volume} {108}},\ \bibinfo
  {pages} {014508} (\bibinfo {year} {2023})}\BibitemShut {NoStop}%
\bibitem [{\citenamefont {Braz}(2024)}]{brazCodeMatrixRPA2024}%
  \BibitemOpen
  \bibfield  {author} {\bibinfo {author} {\bibfnamefont {L.}~\bibnamefont
  {Braz}},\ }\bibfield  {title} {\bibinfo {title} {Code for matrix {{RPA}} in
  magic-angle twisted bilayer graphene},\ }\bibfield  {journal} {\bibinfo
  {journal} {Zenoo}\ }\href {https://doi.org/10.5281/zenodo.10775765}
  {10.5281/zenodo.10775765} (\bibinfo {year} {2024})\BibitemShut {NoStop}%
\bibitem [{\citenamefont {Scalapino}\ \emph {et~al.}(1987)\citenamefont
  {Scalapino}, \citenamefont {Loh},\ and\ \citenamefont
  {Hirsch}}]{scalapinoFermisurfaceInstabilitiesSuperconducting1987a}%
  \BibitemOpen
  \bibfield  {author} {\bibinfo {author} {\bibfnamefont {D.~J.}\ \bibnamefont
  {Scalapino}}, \bibinfo {author} {\bibfnamefont {E.}~\bibnamefont {Loh}},\
  and\ \bibinfo {author} {\bibfnamefont {J.~E.}\ \bibnamefont {Hirsch}},\
  }\bibfield  {title} {\bibinfo {title} {Fermi-surface instabilities and
  superconducting d-wave pairing},\ }\href
  {https://doi.org/10.1103/PhysRevB.35.6694} {\bibfield  {journal} {\bibinfo
  {journal} {Phys. Rev. B}\ }\textbf {\bibinfo {volume} {35}},\ \bibinfo
  {pages} {6694} (\bibinfo {year} {1987})}\BibitemShut {NoStop}%
\bibitem [{\citenamefont {Qi}\ \emph {et~al.}(2008)\citenamefont {Qi},
  \citenamefont {Raghu}, \citenamefont {Liu}, \citenamefont {ino},\ and\
  \citenamefont {Zhang}}]{qiPairingStrengthsTwo2008a}%
  \BibitemOpen
  \bibfield  {author} {\bibinfo {author} {\bibfnamefont {X.-L.}\ \bibnamefont
  {Qi}}, \bibinfo {author} {\bibfnamefont {S.}~\bibnamefont {Raghu}}, \bibinfo
  {author} {\bibfnamefont {C.-X.}\ \bibnamefont {Liu}}, \bibinfo {author}
  {\bibfnamefont {D.~J.}\ \bibnamefont {ino}},\ and\ \bibinfo {author}
  {\bibfnamefont {S.-C.}\ \bibnamefont {Zhang}},\ }\href@noop {} {\bibinfo
  {title} {Pairing strengths for a two orbital model of the {{Fe-pnictides}}}}
  (\bibinfo {year} {2008}),\ \Eprint {https://arxiv.org/abs/0804.4332}
  {arXiv:0804.4332 [cond-mat]} \BibitemShut {NoStop}%
\bibitem [{\citenamefont {Defenu}\ \emph {et~al.}(2023)\citenamefont {Defenu},
  \citenamefont {Donner}, \citenamefont {Macr{\`i}}, \citenamefont {Pagano},
  \citenamefont {Ruffo},\ and\ \citenamefont
  {Trombettoni}}]{defenuLongrangeInteractingQuantum2023}%
  \BibitemOpen
  \bibfield  {author} {\bibinfo {author} {\bibfnamefont {N.}~\bibnamefont
  {Defenu}}, \bibinfo {author} {\bibfnamefont {T.}~\bibnamefont {Donner}},
  \bibinfo {author} {\bibfnamefont {T.}~\bibnamefont {Macr{\`i}}}, \bibinfo
  {author} {\bibfnamefont {G.}~\bibnamefont {Pagano}}, \bibinfo {author}
  {\bibfnamefont {S.}~\bibnamefont {Ruffo}},\ and\ \bibinfo {author}
  {\bibfnamefont {A.}~\bibnamefont {Trombettoni}},\ }\bibfield  {title}
  {\bibinfo {title} {Long-range interacting quantum systems},\ }\href
  {https://doi.org/10.1103/RevModPhys.95.035002} {\bibfield  {journal}
  {\bibinfo  {journal} {Rev. Mod. Phys.}\ }\textbf {\bibinfo {volume} {95}},\
  \bibinfo {pages} {035002} (\bibinfo {year} {2023})}\BibitemShut {NoStop}%
\bibitem [{\citenamefont {Braz}\ \emph {et~al.}(2024)\citenamefont {Braz},
  \citenamefont {Martins},\ and\ \citenamefont {{da
  Silva}}}]{brazChargeSpinFluctuations2024}%
  \BibitemOpen
  \bibfield  {author} {\bibinfo {author} {\bibfnamefont {L.~B.}\ \bibnamefont
  {Braz}}, \bibinfo {author} {\bibfnamefont {G.~B.}\ \bibnamefont {Martins}},\
  and\ \bibinfo {author} {\bibfnamefont {L.~G. G. V.~D.}\ \bibnamefont {{da
  Silva}}},\ }\href@noop {} {\bibinfo {title} {Charge and spin fluctuations in
  superconductors with intersublattice and interorbital interactions}}
  (\bibinfo {year} {2024}),\ \Eprint {https://arxiv.org/abs/2403.02453}
  {arXiv:2403.02453 [cond-mat]} \BibitemShut {NoStop}%
\end{thebibliography}

%

\end{document}